\documentclass[fleqn,usenatbib]{mnras}

\usepackage{newtxtext,newtxmath}

\usepackage[T1]{fontenc}

\DeclareRobustCommand{\VAN}[3]{#2}
\let\VANthebibliography\thebibliography
\def\thebibliography{\DeclareRobustCommand{\VAN}[3]{##3}\VANthebibliography}

\usepackage{graphicx}	
\usepackage{amsmath}	

\usepackage{comment}

\title[Hybrid bispectrum]{Validation of the Hybrid Bias Expansion model for the galaxy bispectrum}

\author[Pellejero Ibáñez et al.]
{Marcos Pellejero Ib\'a\~nez$^{1}$\thanks{E-mail: mpelleje@ed.ac.uk}, Raul E. Angulo$^{2,3}$, Laila Linke$^{4}$, Sara Ortega-Martinez$^{2,5}$, 
Maria Tsedrik$^{1}$, \newauthor
Sergio Contreras$^{6}$, John A. Peacock$^{1}$, Kate Storey-Fisher$^{2,7}$, Jens St\"ucker$^8$, Rodrigo Voivodic$^{2}$,  \newauthor
and Matteo Zennaro$^{9}$
\\
\\
$^{1}$Institute for Astronomy, University of Edinburgh, Royal Observatory, Blackford Hill, Edinburgh, EH9 3HJ, UK\\
$^{2}$Donostia International Physics Center (DIPC), Paseo Manuel de Lardizabal 4, 20018 Donostia-San Sebastian, Spain\\
$^{3}$IKERBASQUE, Basque Foundation for Science, E-48013, Bilbao, Spain\\
$^{4}$Universität Innsbruck, Institut für Astro- und Teilchenphysik, Technikerstr. 25/8, 6020 Innsbruck, Austria \\
$^{5}$University of the Basque Country UPV/EHU, Department of Theoretical Physics, Bilbao, E-48080, Spain\\
$^{6}$Facultad de F\'isica, Universidad de Sevilla, Campus de Reina Mercedes, Av. Reina Mercedes s/n 41012 Seville, Spain \\
$^{7}$Kavli Institute for Particle Astrophysics and Cosmology, Stanford University, 452 Lomita Mall, Stanford, CA 94305, USA \\
$^{8}$University of Vienna, Department of Astrophysics,
Turkenschanzstraße 17, 1180 Vienna, Austria \\
$^{9}$Institute of Space Sciences (ICE, CSIC), Campus UAB, Carrer de Can Magrans, s/n, 08193 Barcelona, Spain
}

\date{Accepted XXX. Received YYY; in original form ZZZ}

\pubyear{\the\year{}}

\begin{document}
\label{firstpage}
\pagerange{\pageref{firstpage}--\pageref{lastpage}}
\maketitle

\begin{abstract}
The Hybrid Bias Expansion model (also known as Hybrid Effective Field Theory, HEFT) provides a promising way to extend the range of validity of perturbative large-scale structure modelling by replacing perturbative gravitational evolution with the nonlinear displacement field measured from $N$-body simulations. While this approach has already been shown to improve the modelling of the power spectrum, its validity at the bispectrum level has not yet been established. In this work we perform a first systematic real-space validation of the Hybrid bispectrum model using DESI-like LRG and ELG mock catalogues constructed at fixed cosmology on volumes similar to those of DESI's LRG samples. We find that the model remains self-consistent up to $k_{\rm max}^B \simeq 0.25\,h\,{\rm Mpc}^{-1}$, while clear signs of breakdown appear for a similar EFT tree-level bispectrum approach at $k_{\rm max}^B \gtrsim 0.13\,h\,{\rm Mpc}^{-1}$. We also show that adding matter cross-statistics significantly improves the precision of the recovered bias parameters, while a partial third-order extension including only the $\delta^3$ operator does not extend the validity range. Finally, we find a strong hierarchy among the bispectrum basis terms when grouped by total bias-operator order, with the lowest-order sectors dominating the total amplitude, which has important implications in emulation strategies.

\end{abstract}




\section{Introduction}

The galaxy distribution is one of the main observables of large-scale structure surveys, and its accurate theoretical modelling is essential for extracting cosmological information from current and forthcoming data sets \citep{4MOST,SPHEREX,miniJPAS, EuclidI_2025, DESIDR1_2026}. In the mildly non-linear regime, perturbative descriptions based on Effective Field Theory (EFT) have become the standard framework for modelling galaxy clustering \citep{Bernardeau_2002,Baumann_2012,Carrasco_2012,Carrasco_2014,Porto_2014,Senatore_2015,Vlah_2015,Mirbabayi_2015,Desjacques2018,Chen_2020,Chen_2021,Schmidt2021,Cabass_2023}. 

These methods provide a controlled expansion of tracer statistics in terms of long-wavelength fields and bias parameters, and have been used successfully in a wide range of cosmological analyses. However, their practical range of validity is limited by the perturbative treatment of gravitational evolution, particularly once one moves beyond the power spectrum and attempts to exploit higher-order statistics such as the bispectrum.

The bispectrum is a particularly valuable observable because it is sensitive to mode coupling, non-Gaussianity generated by gravitational evolution, and the structure of the galaxy bias expansion \citep{Scoccimaro_2000,Philcox_2021,Philcox_2022,Ivanov_2023,Philcox_2024}. In principle, it carries a large amount of information complementary to that of the power spectrum. In practice, however, modelling the bispectrum accurately into the mildly non-linear regime remains challenging. Empirical fitting formulae for the nonlinear matter bispectrum, such as \textsc{BiHalofit}, can reach small scales by calibrating directly against suites of numerical simulations \citep{Takahashi_2020}. These approaches are extremely useful for applications where the matter bispectrum is the relevant quantity, but their accuracy is tied to the simulations and cosmological models used for calibration, and they do not by themselves provide a first-principles description of biased tracers with a controlled bias expansion. By contrast, perturbative EFT descriptions of the galaxy bispectrum provide such an operator-based framework, but typically cease to be reliable at scales around $k \sim 0.15\,h\,{\rm Mpc}^{-1}$ \citep{DAmico_2024}, limiting the fraction of available bispectrum information that can be used robustly.

The Hybrid Bias Expansion model (also known as Hybrid Effective Field Theory, HEFT: \citealt{Modi_2020,PellejeroIbanez2022}) offers a possible route beyond this limitation, and has already been applied in the context of projected clustering and weak-lensing analyses \citep{Hadzhiyska2021, Pellejero-Ibanez_2024b}. The key idea is to retain the perturbative Lagrangian bias expansion, while replacing the perturbative treatment of gravitational evolution with the nonlinear displacement field measured directly from $N$-body simulations. In this way, the complicated mode coupling generated by nonlinear structure formation is incorporated non-perturbatively, while the tracer dependence is still encoded in a finite set of bias coefficients. Previous work has shown that this approach can substantially improve the modelling of the power spectrum \citep{Modi_2020,Kokron2021,ZennaroAnguloPellejero2021,PellejeroIbanez2022,PellejeroIbanez2023}. Whether the same is true at the bispectrum level, however, has remained unclear.

The main obstacle is not conceptual but practical. Extending the Hybrid model to the bispectrum requires the computation of a large number of basis correlators, and a full cosmological application would additionally require emulating their dependence on cosmological parameters \citep{Angulo_2021,baccoemu2023,PellejeroIbanez_2024}. This quickly becomes a demanding computational problem. In the present work we therefore adopt a more focused strategy. Rather than building a cosmology-dependent emulator, we work at fixed cosmology and ask a more basic question: \emph{given the true basis fields of a simulation, up to what scale do we recover the independently measured bias parameters without signs of scale dependence, while providing a good fit to the power spectrum and bispectrum?}

To answer this question, we construct DESI-like LRG and ELG \citep{DESI,Zhou_2023,Raichoor_2023} mock galaxy catalogues from a fixed-cosmology simulation (as presented in \citealt{Contreras_2024, Ortega-Martinez_2025}) and independently measure their Lagrangian bias parameters using the probabilistic bias framework \citep{StueckerPB_2024}. These measurements provide a reference set of large-scale bias coefficients that play the role of `true' field-level values. We then build the basis terms for the power spectrum and bispectrum and fit the clustering statistics of the mock catalogues while varying the maximum bispectrum scale included in the analysis. The central logic of the paper is that the model should be regarded as valid only as long as the fitted clustering parameters remain consistent with the independently measured probabilistic-bias values and do not show any systematic drift with scale.

This setup provides a physically motivated and conservative criterion for identifying the breakdown scale of the model. A fit can remain statistically acceptable even after the theoretical description has begun to fail, because the free parameters may absorb inaccuracies in the model. For cosmological applications, however, such behaviour is not sufficient: the recovered parameters must still correspond to the same underlying large-scale bias coefficients. Note that a simple good fit at fixed cosmology is not enough, since it can be achieved by bias parameters absorbing theory error; this would bias cosmology when cosmology is varied. By comparing clustering-based constraints to independently measured probabilistic-bias values, we can determine where this interpretation ceases to hold.

The paper is organised as follows. In Section~\ref{sec:mocks} we describe the simulations, mock galaxy catalogues, and clustering measurements. In Section~\ref{sec:model} we present the Hybrid Bias Expansion model for the power spectrum and bispectrum in real space. Section~\ref{sec:prob_bias} summarises the probabilistic bias framework and the independent calibration of the bias parameters. In Section~\ref{sec:methodology} we describe the statistical methodology and the criterion used to assess model validity. Section~\ref{sec:results} presents the main results, including the breakdown scale of the Hybrid bispectrum model. In Section~\ref{sec:basis_terms} we discuss the relative importance of the bispectrum basis terms and their implications for future emulator strategies. Finally, Section~\ref{sec:conclusion} summarises our conclusions.

\section{Simulations and Galaxy Mock Catalogues}
\label{sec:mocks}

\subsection{N-body simulations}

Our analysis is based on two gravity-only $N$-body simulations from the BACCO suite \citep{Angulo_2021}, which provide a controlled environment for testing the Hybrid description of tracer clustering in the mildly non-linear regime.

Each simulation evolves $3072^3$ particles in a periodic box of side length $L=1.024\,h^{-1}{\rm Gpc}$, corresponding to a particle mass of $m_p \simeq 3.2\times10^9\,h^{-1}M_\odot$. Gravitational evolution is performed with a modified version of \texttt{L-Gadget3}, a memory-optimised version of the \texttt{Gadget} code family \citep{GADGET2}. The adopted cosmology is consistent with the Planck 2018 $\Lambda$CDM model \citep{PlanckCosmo2020}, with parameters
$\Omega_m = 0.309$,
$\Omega_b = 0.0486$,
$\Omega_\Lambda = 0.6911$,
$h = 0.6774$,
$n_s = 0.9667$, and
$\sigma_8 = 0.8159$.

The box size and redshift are chosen to be representative of the DESI samples that motivate this work \citep{DESIDR1_2026}. For comparison, the DESI DR1 LRG2 sample has an effective volume $V_{\rm eff}\simeq 4.0\,{\rm Gpc}^3$ at $z_{\rm eff}\simeq 0.706$, while the ELG1 sample has $V_{\rm eff}\simeq 2.0\,{\rm Gpc}^3$ at $z_{\rm eff}\simeq 0.955$ \citep{Novel-Masot_2025}. Interpreting these volumes as equivalent cubic volumes and converting with $h\simeq 0.7$ gives effective side lengths of approximately $1.1\,h^{-1}{\rm Gpc}$ and $0.88\,h^{-1}{\rm Gpc}$, respectively. The BACCO box size therefore lies between the effective scales of the two DESI samples and is particularly close to that of the LRG2 bin. We analyse the snapshot at $z=0.8$, which lies between the effective redshifts of these LRG and ELG samples and provides a common redshift at which to construct both DESI-like tracers.

The initial conditions are generated using the paired-and-fixed technique \citep{Angulo2016}. In the first realisation, the amplitudes of the linear modes are fixed to their ensemble-average values, reducing large-scale sample variance. The second simulation uses the same amplitudes but phases shifted by $180^\circ$, forming the paired counterpart. Unless otherwise stated, all measurements shown in this work correspond to the average over the two realisations. This suppresses cosmic variance beyond what would be obtained from two independent simulations of the same volume \citep{Angulo2016,Maion2022}.

To reduce the computational cost of field-level operator measurements, we construct a subsampled matter catalogue by selecting one particle out of every $4^3$ particles in Lagrangian space, leaving $1/64$ of the original particle load. We have verified that this has negligible impact on the large-scale spectra relevant for our analysis. Halo catalogues and halo properties are always measured from the full particle simulation on the fly.

Dark matter haloes are identified with a Friends-of-Friends algorithm \citep{Davis_1985} with linking length $b=0.2$, requiring a minimum of 20 particles per halo.

\subsection{DESI-like LRG mock catalogues}

We construct luminous red galaxy (LRG) mock catalogues using the SHAMe model (\textit{SubHalo Abundance Matching extended}: \citealt{ContrerasAnguloZennaro2020AB, ContrerasAnguloZennaro2020}), which augments standard abundance matching with orphan galaxies, tidal disruption, and flexible assembly-bias contributions.

In SHAMe, galaxy stellar masses are assigned through a parametrised relation controlled by five parameters, $\{\sigma_{M_\star}, t_{\rm merger}, f_c, f_s, \beta_{\rm lum}\}$, which govern the scatter in the abundance-matching relation, the treatment of merger and disruption processes, assembly-bias contributions and, satellite survival. For the DESI-like LRG sample considered here we adopt the parameter set obtained by \cite{Contreras_2024} to best fit the galaxy clustering of the FLAMINGO \citep{Schaye_2023,Kugel_2023,Helly_2026} galaxy sample. We refer to the original paper for the details of the mock production. Note that producing a mock that matches DESI LRG clustering exactly is beyond the scope of this paper; but what we use here is still a realistic mock.

The number density $\bar n \simeq 8\times10^{-4}\,(h^{-1}{\rm Mpc})^{-3}$ is also close to that expected for the DESI LRG sample motivating this work. Using the DR1 LRG2 object count quoted in \citet{Novel-Masot_2025}, $N_{\rm LRG2}=771,875$, together with its effective volume $V_{\rm eff}\simeq 4.0\,{\rm Gpc}^3$, gives an approximate effective number density $\bar n_{\rm LRG2} \simeq 5.6\times10^{-4}\,(h^{-1}{\rm Mpc})^{-3}$, where we have used $h\simeq 0.7$ for this estimate. Our choice, $\bar n \simeq 8\times10^{-4}\,(h^{-1}{\rm Mpc})^{-3}$, is therefore comparable to the DESI DR1 LRG2 effective density and slightly higher, as expected for future samples with increased statistics.

\subsection{DESI-like ELG mock catalogues}

We generate emission-line galaxy (ELG) mock catalogues using the SHAMe-SF model (\textit{SubHalo Abundance Matching extended -- Star Forming}, \citealt{Ortega-Martinez_2024,Ortega-Martinez_2025}), an extension of abundance matching tailored to star-forming populations.

In this framework, subhaloes are rank-ordered according to a modelled star-formation rate parametrised in terms of the peak circular velocity ($V_{\rm peak}$, the maximum over redshift of the maximum circular velocity) and a concentration proxy, modulated by a quenching mechanism dependent on host halo mass and the time since each subhalo reached its peak mass. Once ranked, the final galaxy selection is made using a cut based on the target sample's number density. The free parameters of the model, regulating the functional form, scatter and quenching strength, were calibrated to reproduce the clustering of the ELG sample with $0.8<z<1.1$ of the DESI One-Percent data release \citep{DESI_EDR_2023} from \cite{Rocher_2023}.

The target number density is chosen to be representative of DESI-like ELG samples. Using the DR1 ELG1 object count quoted in \citet{Novel-Masot_2025}, $N_{\rm ELG1}=1,016,340$, together with its effective volume $V_{\rm eff}\simeq 2.0\,{\rm Gpc}^3$, gives the approximate effective number density
$\bar n_{\rm ELG1}
\simeq
1.5\times10^{-3}\,(h^{-1}{\rm Mpc})^{-3}$,
again using $h\simeq 0.7$. We adopt $\bar n=10^{-3}\,(h^{-1}{\rm Mpc})^{-3}$, which is of the same order as the DESI DR1 ELG1 effective density while remaining slightly more conservative.

The resulting catalogues reproduce the clustering properties expected for star-forming tracers as shown in \cite{Ortega-Martinez_2025} while maintaining a physically motivated connection to the underlying halo population.

\subsection{Measured clustering statistics}

From each mock catalogue we measure the real-space two-point and three-point statistics used throughout this work:
\begin{itemize}
    \item the galaxy auto-power spectrum, $P_{gg}(k)$,
    \item the galaxy--matter cross-power spectrum, $P_{gm}(k)$,
    \item the galaxy auto-bispectrum, $B_{ggg}(k_1,k_2,k_3)$,
    \item the galaxy--galaxy--matter bispectrum, $B_{ggm}(k_1,k_2,k_3)$.
\end{itemize}

The matter field entering $P_{gm}$ and $B_{ggm}$ is measured directly from the simulation particle distribution. Power spectra are measured from Fourier transforms of the density fields using standard estimators.

Bispectra are measured with the GPU-based code \texttt{BIG} (\textit{Bispectrum estimator on GPUs}:  \citealt{BiG, Burger_2026}), which implements an FFT-based estimator. Rather than explicitly enumerating all Fourier-space mode triplets satisfying the triangle condition, the estimator exploits the Fourier representation of the Dirac delta enforcing closure and recasts the measurement into a sequence of FFTs, shell-filtered fields, real-space multiplications, and volume averages. This greatly reduces the computational cost of measuring the bispectrum over the full set of triangle configurations relevant for our analysis. However, the cost is still high, while the power spectra only take a few tens of seconds of CPU time on a 768$^3$ grid, the evaluation of one bispectrum takes around 1.5 GPU hours for 20 bins between $k_{\rm min}=0.01227\,h\,{\rm Mpc}^{-1}$ and $k_{\rm max}=0.4\,h\,{\rm Mpc}^{-1}$ (giving a total of 8,000 bins as we explain in the next section). 

Because the simulations are fixed and paired, the variance of the measured large-scale spectra is reduced relative to that of a standard ensemble of independent realisations. We return to the implications of this point for the covariance model in Section~\ref{sec:methodology}.

\section{Hybrid EFT Modelling of Power Spectrum and Bispectrum}
\label{sec:model}

\subsection{The Hybrid Bias Expansion in real space}

The goal of the Hybrid Bias Expansion (also known as Hybrid Effective Field Theory, HEFT) model is to describe biased tracer clustering by combining a perturbative bias expansion in the initial conditions with a non-perturbative treatment of gravitational evolution \citep{Modi_2020,PellejeroIbanez2022}. In this work we restrict the analysis to real space.

As in Lagrangian descriptions of structure formation \citep{Zeldovich_1970,Matsubara2008,White_2014,Desjacques2018}, each mass element is labelled by its initial coordinate $\mathbf q$, and its Eulerian position at late time is written as
\begin{equation}
\mathbf x = \mathbf q + \boldsymbol{\psi}(\mathbf q),
\end{equation}
where $\boldsymbol{\psi}(\mathbf q)$ is the displacement field. In standard perturbative EFT approaches this displacement is itself expanded perturbatively. In the Hybrid approach, by contrast, $\boldsymbol{\psi}$ is measured directly from the $N$-body simulation. This is the key ingredient that allows the method to retain the full nonlinear gravitational evolution of the matter field while keeping a controlled, low-order description of tracer bias.

For matter, mass conservation implies
\begin{equation}
1+\delta_m(\mathbf x) =
\int d^3q\,
\delta_{\rm D}\!\left(\mathbf x-\mathbf q-\boldsymbol{\psi}(\mathbf q)\right),
\end{equation}
so that the evolved matter density field is fully determined by the displacement field. In other words, the nonlinear matter field is obtained by advecting initially uniform mass elements under the measured displacements.

The tracer field is constructed in the same spirit, except that different Lagrangian environments are assigned different weights before advection. This is the point at which bias enters the Hybrid framework.

\subsection{Lagrangian bias as environment-dependent weighting}

Galaxies or haloes do not form with equal probability in all regions of the initial density field. Instead, their abundance depends on the local large-scale environment, which in the Lagrangian picture can be described through a bias expansion in operators constructed from the initial fields \citep{Assassi2014,Desjacques2018,Schmidt2021}. In the Hybrid model, this dependence is encoded through a Lagrangian weighting function $w(\mathbf q)$, which assigns a different weight to each initial fluid element according to a set of operators built from the linear density field.

The tracer density field is then obtained by advecting these weighted fluid elements with the same displacement field that moves the dark matter \citep{Modi_2020}:
\begin{equation}
1+\delta_g(\mathbf x) =
\int d^3q\, w(\mathbf q)\,
\delta_{\rm D}\!\left(\mathbf x-\mathbf q-\boldsymbol{\psi}(\mathbf q)\right).
\end{equation}
Relative to the matter case, the only difference is the factor $w(\mathbf q)$. Regions of the initial conditions with larger tracer weight contribute more strongly to the final tracer density, while underweighted environments contribute less. The final clustering therefore results from the combination of two ingredients: the nonlinear advection encoded in $\boldsymbol{\psi}$ and the initial environment dependence encoded in $w$.

Up to second order, we model the weight as \citep{Assassi2014,Desjacques2018}
\begin{equation}
\begin{aligned}
w(\mathbf q) = &\,1 + b_1\,\delta_{\rm L}(\mathbf q) + b_2\left[\delta_{\rm L}^2(\mathbf q)-\langle \delta_{\rm L}^2\rangle\right] \\ &+ b_{s^2}\left[s^2(\mathbf q)-\langle s^2\rangle\right] + b_{\nabla^2\delta}\,\nabla^2\delta_{\rm L}(\mathbf q).
\end{aligned}
\end{equation}
Here $\delta_{\rm L}$ is the linear density field, and
\begin{equation}
s^2=s_{ij}s_{ij},
\qquad s_{ij}(\mathbf q)= \left(\frac{\partial_i\partial_j}{\nabla^2}-\frac{1}{3}\delta^{\rm K}_{ij}\right)\delta_{\rm L}(\mathbf q),
\end{equation}
with $\delta^{\rm K}_{ij}$ the Kronecker delta. The $b_i$ parameters are the so called bias parameters, assumed to be independent of scale. The subtraction of expectation values for the quadratic operators ensures that they have zero mean, so that the corresponding coefficients can be interpreted as renormalised large-scale bias parameters. The subtraction of expectation values should not be confused with a full statistical decorrelation of the operators. It fixes the zero-point of the composite fields and removes contributions degenerate with lower-order operators in the large-scale bias expansion, but the different environmental variables remain correlated in the initial conditions.

It is convenient to define the renormalised operator basis 
\begin{align}
        \mathcal O_0(\mathbf q) &=1, \qquad
        \mathcal O_1(\mathbf q)=\delta_{\rm L}, \qquad
        \mathcal O_2(\mathbf q)=\delta_{\rm L}^2-\langle \delta_{\rm L}^2\rangle, \qquad \\ &
        \mathcal O_3(\mathbf q)=s^2-\langle s^2\rangle, \qquad
        \mathcal O_4(\mathbf q)=\nabla^2\delta_{\rm L}. \nonumber
      \end{align}
In terms of this basis, the weight can be written compactly as
\begin{equation}
w(\mathbf q)=
\sum_{i=0}^4\,\mathcal O_i(\mathbf q) \qquad b_0=1,
\end{equation}
so that in effect the constant $O_0$ operator is included in the full sum, but with coefficient fixed to unity. This ensures that the weight field has an expectation value of unity. The tracer density field then becomes
\begin{equation}
1+\delta_g(\mathbf x)
=
\sum_i b_i
\int d^3q\, \mathcal O_i(\mathbf q)\,
\delta_{\rm D}\!\left(\mathbf x-\mathbf q-\boldsymbol{\psi}(\mathbf q)\right).
\end{equation}

This expression shows explicitly that each operator defines its own advected contribution to the final tracer field. A central assumption of the model is that the coefficients $b_i$ are constants, independent of scale. Testing the scale at which this assumption ceases to hold is one of the main goals of this paper.

\subsection{Advected basis fields}

Motivated by the previous expression, and following the hybrid-bias construction of previous applications \citep{Modi_2020,Kokron2021,ZennaroAnguloPellejero2021,PellejeroIbanez2022,PellejeroIbanez2023,PellejeroIbanez_2024}, we define a set of Eulerian basis fields by advecting each Lagrangian operator separately,
\begin{equation}
\delta_i(\mathbf x)
\equiv
\int d^3q\, \mathcal O_i(\mathbf q)\,
\delta_{\rm D}\!\left(\mathbf x-\mathbf q-\boldsymbol{\psi}(\mathbf q)\right).
\end{equation}
For the constant operator, this reduces to the matter overdensity, $\delta_0=\delta_m$. The tracer overdensity can then be written as the linear combination
\begin{equation}
\delta_g(\mathbf x)=\sum_i b_i\,\delta_i(\mathbf x).
\end{equation}
where we have ignored any stochastic terms coming from the point-like nature of galaxies.
This decomposition is the practical basis of the Hybrid model. Once the fields $\delta_i$ are measured from the simulation, any power spectrum or bispectrum prediction can be assembled as a linear combination of correlators among these fields, with coefficients determined by the bias parameters.

The advantage of this formulation is that the complicated nonlinear gravitational evolution is already encoded in the measured basis fields. The fitting problem is therefore reduced to determining the constant coefficients multiplying a finite set of precomputed correlators.

\subsection{Power-spectrum model}

The galaxy auto-power spectrum is defined by
\begin{equation}
\langle \delta_g(\mathbf k)\delta_g(\mathbf k')\rangle
=
(2\pi)^3\delta_{\rm D}(\mathbf k+\mathbf k')\,P_{gg}(k).
\end{equation}
Substituting the basis expansion for $\delta_g$, we obtain
\begin{equation}
P_{gg}(k)=\sum_{ij} b_i b_j\,P_{ij}(k),
\end{equation}
which corresponds to the standard power-spectrum implementation of the hybrid Lagrangian-bias model \citep{ZennaroAnguloPellejero2021}, where
\begin{equation}
P_{ij}(k)\equiv
\langle \delta_i(\mathbf k)\delta_j(-\mathbf k)\rangle
\end{equation}
denotes the auto- or cross-power spectrum of the advected basis fields. 
Since the constant operator corresponds to the matter field, the mixed galaxy--matter spectrum is obtained from the same basis as
\begin{equation}
P_{gm}(k)=\sum_i b_i\,P_{i0}(k).
\end{equation}

The stochastic contribution to the tracer auto-power spectrum is modelled through a low-$k$ expansion as 
\begin{equation}
P_{\rm noise}(k)
=
\frac{1}{\bar n_g}
\left(
\epsilon_{P,1}+\epsilon_{P,2}k^2
\right)
\end{equation}
\citep{Perko_2016},
where $\bar n_g$ is the tracer number density, and the noise is included as an additive contribution to the spectrum:
\begin{equation}
P_{gg}(k)=\sum_{ij} b_i b_j\,P_{ij}(k)+P_{\rm noise}(k).
\end{equation}
Note that here we rename $P_{gg}$ as the galaxy spectrum including stochasticity, and will do so for the rest of the paper.
The constant term $\epsilon_{P,1}$ captures non-Poisson constant stochasticity from exclusion, satellite occupation, halo occupation scatter, and nontrivial galaxy sampling, while the $\epsilon_{P,2}$ term accounts for leading scale-dependent stochasticity.

Because we use five basis operators, the power-spectrum model involves all pairs of operators $(i,j)$. If one keeps the ordering of the fields, there are $5\times5=25$ possible products. However, since the power spectrum is symmetric under interchange of the two fields, $P_{ij}(k)=P_{ji}(k)$, only the unique unordered pairs need to be computed. The number of independent power-spectrum basis terms is therefore
$N_P=5(5+1)/2=15$.

\subsection{Bispectrum model}

We extend the same basis-field construction to the bispectrum, whose perturbative modelling has been developed and applied in several recent EFT analyses \citep{Philcox_2021,Philcox_2022,Ivanov_2023,DAmico_2024,Philcox_2024}. The galaxy bispectrum is defined by
\begin{equation}
\langle
\delta_g(\mathbf k_1)\delta_g(\mathbf k_2)\delta_g(\mathbf k_3)
\rangle
=
(2\pi)^3
\delta_{\rm D}(\mathbf k_1+\mathbf k_2+\mathbf k_3)\,
B_{ggg}(k_1,k_2,k_3).
\end{equation}
Using the basis expansion of the tracer field, the deterministic prediction becomes
\begin{equation}
B_{ggg}(k_1,k_2,k_3)
=
\sum_{ijk}
b_i b_j b_k\,
B_{ijk}(k_1,k_2,k_3)
+
B_{\rm noise}^{ggg}(k_1,k_2,k_3),
\end{equation}
where
\begin{equation}
B_{ijk}(k_1,k_2,k_3)
\equiv
\langle
\delta_i(\mathbf k_1)\delta_j(\mathbf k_2)\delta_k(\mathbf k_3)
\rangle.
\end{equation}
Again identifying the constant operator with the matter field, the mixed bispectrum can be written as
\begin{equation}
B_{ggm}(k_1,k_2,k_3)
=
\sum_{ij}
b_i b_j\,
B_{ij0}(k_1,k_2,k_3)
+
B^{ggm}_{\rm noise}(k_1,k_2,k_3).
\label{eq:Bggm}
\end{equation}
One could analogously define the bispectrum with one galaxy leg and two matter legs,
\begin{equation}
B_{gmm}(k_1,k_2,k_3)=\sum_i b_i\,B_{i00}(k_1,k_2,k_3)+B^{gmm}_{\rm noise}(k_1,k_2,k_3),
\end{equation}
where the index $0$ denotes the constant operator, i.e. the matter field. We do not include this statistic in the fiducial analysis. Our aim is to use matter cross-statistics to help break bias degeneracies while retaining strong sensitivity to nonlinear galaxy-bias combinations. For this purpose, $B_{ggm}$ is the closest matter-cross analogue of $B_{ggg}$, since it still depends quadratically on the bias parameters, whereas $B_{gmm}$ depends only linearly on them. We return to this choice in Section~\ref{sec:methodology}.

The bispectrum is where the extension of the model becomes substantially more demanding than for the power spectrum. With the same five operators, one now needs all triplets $(i,j,k)$. Since in our implementation each operator can appear on a different leg of the triangle, the relevant object is an ordered triplet rather than an unordered set. The number of deterministic basis terms is therefore
$N_B=5^3=125$.

This difference with the power spectrum is important. For the two-point function, exchanging the two legs does not generate a new basis term because the correlator is symmetric. For the bispectrum, by contrast, the three legs are associated with $(k_1,k_2,k_3)$, and the contributions of different operator assignments must be tracked explicitly when assembling the model over all triangle configurations. Even when some permutations are related by symmetry, the bookkeeping and measurement cost scale as the full cube of the number of operators.

This rapid combinatorial growth is one of the main reasons why bispectrum-level Hybrid theory has not yet been explored as systematically as the power spectrum. It is also the reason why constructing a general cosmology emulator for all bispectrum basis terms is computationally challenging and beyond the scope of this paper.

\subsection{Stochastic contributions to the bispectrum}

In the bispectrum case, besides a constant shot-noise term, one expects contributions coupling stochastic fields to long-wavelength clustering, leading to terms proportional to power spectra and to low-order powers of the triangle sides \citep{Desjacques2018,Schmidt2021,Cabass_2023,DAmico_2024}.

Motivated by the next-to-leading-order stochastic contribution to the real-space galaxy bispectrum in biased-tracer EFT \citep{Eggemeier_2019,Philcox1loop_2022}, we adopt the phenomenological form
\begin{equation}
\begin{aligned}
B_{\rm noise}^{ggg}(k_1,k_2,k_3) = & \frac{1}{\bar n_g^2} \left[ 1+\epsilon_{B,0} +\epsilon_{B,1}(k_1^2+k_2^2+k_3^2) \right] \\ & + \frac{1}{\bar n_g} \big[ (1+\epsilon_{B,2})(P_1+P_2+P_3) \\ & + \epsilon_{B,3}\big(P_1(k_2^2+k_3^2)+{\rm cyc.}\big) +\epsilon_{B,4}\big(P_1k_1^2+{\rm cyc.}\big) \big],
\end{aligned}
\end{equation}
where $P_i\equiv P_{gg}(k_i)$, and `cyc.' denotes cyclic permutations over the triangle legs.

This expression includes the leading constant and scale-dependent shot-noise terms, as well as the lowest-order couplings between stochasticity and large-scale clustering. In practice, these terms absorb residual contributions not captured by the deterministic basis expansion.

For the mixed bispectrum $B_{ggm}$, only the tracer legs contribute to stochasticity, and we model them as
\begin{equation}
\begin{aligned}
B^{ggm}_{\rm noise}(k_1,k_2,k_3)
=
\frac{1}{\bar n_g}
\big[
\eta_0 P_{gm}(k_3)
+ & \eta_1 k_3^2 P_{gm}(k_3) \\ &
+\eta_2 (k_1^2+k_2^2)P_{gm}(k_3)
\big].
\end{aligned}
\end{equation}
This form captures the leading stochastic terms consistent with the structure of the mixed bispectrum.

\subsection{Basis measurements used in this work}

The Lagrangian operators are evaluated from the linear density field after applying a sharp-$k$ filter with cutoff
$k_s = 0.75\,h\,{\rm Mpc}^{-1}$, following the basis-field construction used in previous applications \citep{ZennaroAnguloPellejero2021,PellejeroIbanez2022,PellejeroIbanez2023}.

For the five operators in the fiducial second-order basis, this procedure yields 15 distinct power-spectrum basis terms and 125 bispectrum basis terms. These basis correlators are the basic ingredients from which the model predictions for $P_{gg}$, $P_{gm}$, $B_{ggg}$, and $B_{ggm}$ are assembled.

An important practical point is that, in a future cosmological application, these basis terms would themselves have to be predicted as functions of cosmology, for example through simulation-based emulators \citep{Angulo_2021,baccoemu2023,PellejeroIbanez_2024}.At the power-spectrum level this is already nontrivial; at the bispectrum level the required number of basis terms becomes much larger. This is one of the central computational challenges motivating the strategy adopted in this paper.

Here, instead of emulating the cosmology dependence of the full basis, we work at fixed cosmology and ask a more focused question: given the true Hybrid model basis fields of a simulation, up to what bispectrum scale can the tracer clustering be described with scale-independent bias parameters?

\subsection{Parameter set of the fiducial model}

The deterministic part of the fiducial second-order Hybrid model is specified by the four free bias parameters
\begin{equation}
\{b_1,\;b_2,\;b_{s^2},\;b_{\nabla^2\delta}\}.
\end{equation}
These are supplemented by nuisance parameters describing stochastic contributions to the power spectrum and bispectra.

In dedicated tests we also extend the deterministic basis by including the renormalised cubic operator $[\delta_{\rm L}^3]$, with coefficient $b_3$, in order to assess whether a partial third-order extension improves the range of validity of the model. Following the standard renormalised-bias construction, higher-order operators should be orthogonalised against lower-order operators so that their coefficients do not simply renormalise lower-order bias parameters \citep{McDonald2006,Assassi2014,Desjacques2018,Schmidt2021}. For the cubic density operator, and assuming Gaussian initial conditions, this corresponds to the normal-ordered combination
\begin{equation}
[\delta_{\rm L}^3](\mathbf q)
\equiv
\delta_{\rm L}^3(\mathbf q)
-
3\langle \delta_{\rm L}^2\rangle\,\delta_{\rm L}(\mathbf q),
\end{equation}
where the variance is computed from the same smoothed linear density field used to construct the basis. This subtraction removes the lower-order contribution of the cubic operator that is proportional to the linear density field. Indeed, for Gaussian initial conditions Wick's theorem gives
\begin{equation}
\langle \delta_{\rm L}^3(\mathbf q)\delta_{\rm L}(\mathbf q')\rangle
=
3\langle \delta_{\rm L}^2\rangle
\langle \delta_{\rm L}(\mathbf q)\delta_{\rm L}(\mathbf q')\rangle,
\end{equation}
so that $\smash{[\delta_{\rm L}^3]}$ is orthogonal to $\delta_{\rm L}$ at leading order. This is the cubic analogue of subtracting $\langle\delta_{\rm L}^2\rangle$ from the quadratic density operator, and ensures that the coefficient $b_3$ captures genuinely cubic bias rather than a renormalization of $b_1$. We stress that this is only the density-only cubic contribution. A complete third-order bias expansion would also contain tidal and derivative operators, together with their corresponding renormalised combinations \citep{Desjacques2018,Schmidt2021}.

The same consistency test can in principle be applied to any clustering statistic: if the model is valid, the deterministic bias parameters inferred from progressively smaller scales should remain stable and should correspond to the same large-scale coefficients. At the power-spectrum level, previous applications of the Hybrid bias expansion have shown that this condition can be satisfied to substantially smaller scales than those accessible to perturbative approaches, and in the present analysis we find no evidence that the power spectrum is the limiting statistic over the range used here (e.g. \citealt{ZennaroAnguloPellejero2021}). We therefore fix the power-spectrum cut to $k_{\rm max}^P=0.4\,h\,{\rm Mpc}^{-1}$ in the main Hybrid analysis and use the bispectrum scale $k_{\rm max}^B$ as the variable that tests the previously unexplored regime.

\section{Bias Parameters from the Probabilistic Bias Framework}
\label{sec:prob_bias}

In this work, the bias coefficients are not treated as unconstrained nuisance parameters. Instead, we adopt independent measurements obtained with the probabilistic bias framework \citep{StueckerGPB_2024,StueckerPB_2024}, which provides field-level estimates of the large-scale Lagrangian bias parameters, $b_i$. These measurements play a central role in our analysis: they define the reference values against which the clustering-based constraints are compared in order to assess the scale of validity of the model.

We summarise here only the main ingredients of the method needed for the present paper. A full derivation and detailed discussion of the probabilistic estimators can be found in \citet{StueckerPB_2024}, while the near-Gaussian structure of the Lagrangian bias function is discussed in \citet{StueckerGPB_2024}.

The probabilistic bias framework is rooted in the peak-background split (PBS) interpretation of bias, in which tracer abundance responds to long-wavelength perturbations of the initial conditions \citep{Kaiser1984,Bardeen1986,Mo&White2002,Desjacques2018}. The response of the galaxy density to a contrast in the matter density at infinitely large scales can be written as the one-dimensional function
\begin{equation}
F(\delta_0)\equiv \frac{n_g(\delta_0)}{n_{g,0}}
=
1+b_1^{\rm L}\delta_0+\frac12 b_2^{\rm L}\delta_0^2+\cdots,
\end{equation}
where $\delta_0$ denotes a long-wavelength background perturbation. The corresponding bias coefficients are given by derivatives of this response evaluated at $\delta_0=0$.

Operationally, these coefficients can be inferred directly from simulations by comparing the unconditional distribution of the smoothed linear density field, $p(\delta)$, with the same field sampled at galaxy Lagrangian positions, $p(\delta|g)$ \citep{StueckerGPB_2024,StueckerPB_2024}. In the density-only case, the bias coefficients can be written as
\begin{equation}
b_n = (-1)^n \int d\delta\, \frac{p^{(n)}(\delta)}{p(\delta)}\,p(\delta|g),
\end{equation}
where $p^{(n)}$ denotes the $n$-th derivative of the unconditional distribution with respect to $\delta$. This expression makes explicit that the large-scale bias parameters are encoded in the difference between the environmental distribution seen by tracers and that of the full matter field.

Once additional operators such as the tidal field or higher-derivative terms are included, the problem becomes intrinsically multivariate \citep{Desjacques2018,Schmidt2021,StueckerPB_2024}. Although the operators entering the Hybrid basis are renormalised by subtracting expectation values, and in some cases lower-order pieces, this does not make the underlying environmental variables statistically independent. In particular, at finite smoothing scale the density, its Laplacian, and the components of the tidal tensor have a non-trivial joint distribution. They therefore cannot in general be treated through independent one-dimensional responses. Instead, the tracer response must be described as a function of the full local environment,
\begin{equation}
F = F(\delta,\nabla^2\delta,s^2,\ldots),
\end{equation}
or, equivalently, through the joint conditional distribution of these variables evaluated at tracer Lagrangian positions.

In practice, this implies that a density-only model corresponds to a one-dimensional response, adding the Laplacian requires at least a two-dimensional response $F(\delta,\nabla^2\delta)$, while including tidal terms requires a multivariate response involving the components of the tidal tensor, or an equivalent isotropic tensor decomposition. The scalar coefficients entering the expansion, such as $b_1$, $b_2$, $b_{s^2}$, and $b_{\nabla^2\delta}$, are then obtained as derivatives or tensor projections of this multivariate response evaluated around vanishing background perturbations.

This construction allows one to infer the large-scale bias parameters directly from field-level statistics, without relying on correlation-function measurements. In particular, it provides an estimate of the same large-scale bias coefficients that should be recovered by the fits if the model remains valid over the scales included in the analysis.

However, note an important difference. For the probabilistic-bias measurements we use the tensorial estimators of \citet{StueckerPB_2024} with spatial corrections up to second order. In this setup the bias parameters are measured for the operators $\{J_2,J_{22},J_{2=2},J_4\}$, corresponding respectively to the linear density $\delta$, the quadratic density operator $\delta^2$, the tidal invariant $s^2=s_{ij}s^{ij}$, and the higher-derivative operator $L=\nabla^2\delta$. The spatial-order-two choice does not introduce all second-order operators as independent bias parameters. Instead, the additional fields $J_{2=4}$ and $J_{4=4}$ are included in the estimator in order to remove the leading smoothing-scale dependence of the lower-order bias measurements. These additional spatial-order-two fields entering the estimator can be written in terms of the trace-free tidal tensor $s_{ij}=\partial_i\partial_j\phi-\delta_{ij}\delta/3$ and the trace-free density Hessian $Q_{ij}=\partial_i\partial_j\delta-\delta_{ij}\nabla^2\delta/3$. In this notation, 
\begin{equation}
J_{2=4}=s_{ij}Q_{ij} = \partial_i\partial_j\phi\,\partial_i\partial_j\delta -\frac{1}{3}\delta\nabla^2\delta , 
\end{equation}  
and
\begin{equation}
J_{4=4}=Q_{ij}Q_{ij}
=
(\partial_i\partial_j\delta)^2
-\frac{1}{3}(\nabla^2\delta)^2 .
\end{equation}
These operators are not included as independent bias terms in our fiducial basis, but enter the probabilistic-bias estimator as spatial-order-two corrections. Thus, the probabilistic-bias measurements provide estimates of the same lower-order coefficients $\{b_1,b_2,b_{s^2},b_{\nabla^2\delta}\}$ that enter the fiducial basis, while using $J_{2=4}$ and $J_{4=4}$ as additional spatial-order-two corrections to reduce residual smoothing-scale dependence.

The PBS interpretation requires the linear density field to be smoothed on sufficiently large scales such that the separation between long- and short-wavelength modes is well defined \citep{Kaiser1984,Peacock&Heavens1985,Bardeen1986,Coles1986,Desjacques2018,StueckerPB_2024}. In this work, we evaluate the linear density field using a sharp-$k$ filter with smoothing scale $k_s^{\rm PB}$, chosen to minimise residual scale dependence while maintaining high statistical precision at $z=0.8$.

\begin{figure}
    \includegraphics[width=\columnwidth]{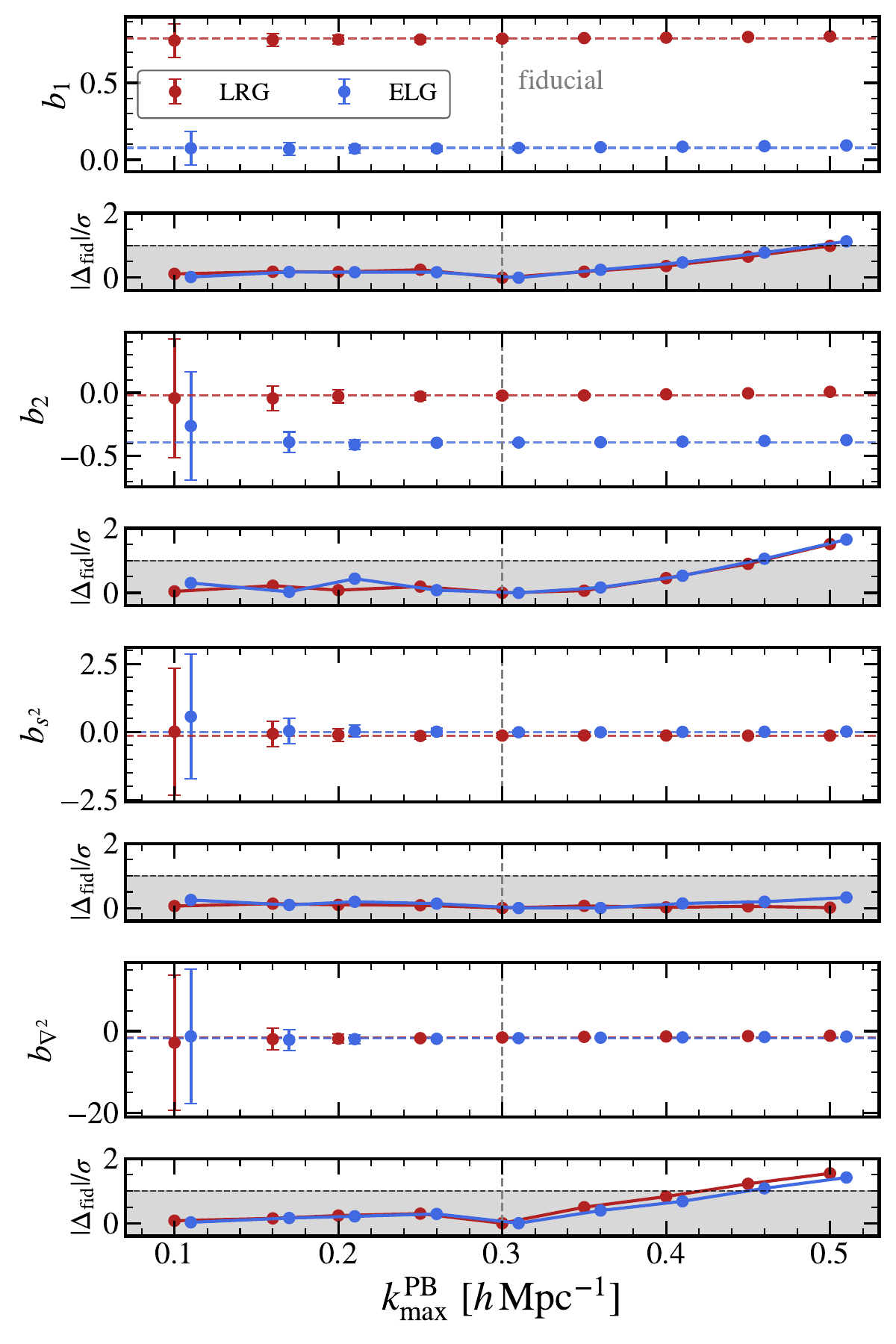}
    \caption{Probabilistic bias measurements of the bias parameters as a function of smoothing scale. Large panels indicate the absolute values while smaller panels show their statistical deviations with respect to the fiducial value chosen in this work. Dashed lines indicate the fiducial values, corresponding to $k_s^{\rm PB}=0.3\,h\,\mathrm{Mpc}^{-1}$. We could have pushed the scale cut to $k_s^{\rm PB}=0.4\,h\,\mathrm{Mpc}^{-1}$ but we choose the previous conservative value that gives deviations smaller than $0.5\sigma$. The ELG results have been shifted to avoid clutter. }
    \label{fig:pb_bias}
\end{figure}

The resulting measurements are shown in Fig.~\ref{fig:pb_bias}. No significant scale dependence is observed down to $k_s^{\rm PB}=0.3\,h\,\mathrm{Mpc}^{-1}$ for the samples considered here. Note that every measurement is consistent within $1 \sigma$ down to scales of $k_s^{\rm PB}=0.4\,h\,\mathrm{Mpc}^{-1}$ but we choose to be extra-conservative here and admit deviations of just a fraction of a $\sigma$.The breaking-scale is very similar in both the ELG and the LRG samples, even though the bias parameters show very different values. We therefore adopt $k_s^{\rm PB}=0.3\,h\,\mathrm{Mpc}^{-1}$ as our reference probabilistic bias values.

Uncertainties are estimated with a jackknife procedure based on 64 sub-volumes of the simulation box. Although approximate, this provides a useful estimate of the statistical precision of the probabilistic-bias measurements. A more robust covariance calibration based on a large ensemble of independent mock catalogues is left for future work.

\section{Statistical Methodology}
\label{sec:methodology}

The objective of the statistical analysis is to determine whether the Hybrid model described in Section~\ref{sec:model} can recover the independently calibrated probabilistic bias parameters of Section~\ref{sec:prob_bias} when fitted to the clustering statistics measured from the mock catalogues.

To this end, we compare its predictions for the power spectrum and bispectrum to the measurements obtained from the simulations, and infer the posterior distribution of the model parameters using a Gaussian likelihood. The analysis is repeated while progressively increasing the maximum bispectrum scale included in the fit. This allows us to identify the point at which the recovered bias parameters become unstable or inconsistent with the probabilistic-bias benchmark.

Throughout this work, the parameters of primary interest are the deterministic bias coefficients,
$\{b_1,b_2,b_{s^2},b_{\nabla^2\delta}\}$,
which we interpret as the same renormalised large-scale Lagrangian bias parameters measured independently with the probabilistic-bias framework \citep{StueckerPB_2024}. Stochastic contributions are treated as nuisance parameters and marginalised over.

\subsection{Data vectors}

We consider several combinations of clustering observables in order to isolate the impact of the matter cross-statistics. The full set of measured quantities is
\begin{equation}
\mathbf D=
\{P_{gg},\,P_{gm},\,B_{ggg},\,B_{ggm}\},
\end{equation}
where only the subset relevant to a given fit is included. The matter cross-statistic included at the bispectrum level is $B_{ggm}$ rather than $B_{gmm}$. This choice is motivated by the goal of the present validation test. Since we are primarily interested in the stability of the galaxy-bias expansion, it is useful to include a matter leg while retaining strong sensitivity to the nonlinear galaxy-bias parameters. Thus, $B_{ggm}$ is the closest matter-cross analogue of the galaxy bispectrum: it contains one unbiased matter leg, which helps to break degeneracies, while still depending quadratically on the galaxy-bias coefficients as shown in Eq.\ref{eq:Bggm}. By contrast, $B_{gmm}$ contains only one biased tracer leg and is therefore less directly sensitive to the nonlinear bias combinations that are central to our breakdown criterion. Moreover, the projected counterpart of $B_{ggm}$, involving two galaxy-density fields and one lensing convergence field, is a natural extension of galaxy-galaxy lensing to three-point statistics. We therefore use $B_{ggm}$ as the minimal and most relevant matter-cross bispectrum for the purposes of this work. Including $B_{gmm}$ would provide additional information, but would also enlarge the data vector and covariance without changing the validation question addressed here.

For the power spectra, Fourier bins are included up to a fixed maximum wavenumber
$k_{\max}^{P}=0.4\,h\,{\rm Mpc}^{-1}$.
For the bispectra, all triangle configurations satisfying
$k_i \le k_{\max}^{B}$
are included, where $k_{\max}^{B}$ is varied throughout the analysis.

\subsection{Gaussian covariance approximation}

For parameter inference we use an analytic covariance evaluated in the Gaussian-field approximation. That is, when computing the covariance of the power-spectrum and bispectrum estimators, we assume that the relevant density fields have Gaussian statistics and evaluate the required higher-order moments using Wick contractions, \citet{Chan&Blot2017}. Under this approximation, different Fourier modes are independent, and the covariance is diagonal in Fourier-bin space for the power spectrum and in triangle-bin space for the bispectrum. For the joint fits to $\{P_{gg},P_{gm}\}$ and to $\{B_{ggg},B_{ggm}\}$, we retain the Gaussian cross-covariances between the two statistics in each pair. In the combined power-spectrum plus bispectrum fits, we neglect the cross-covariance between the power-spectrum and bispectrum sectors. In the Gaussian approximation this mixed covariance is absent for zero-mean fields, since it involves an odd-order moment, while in the true non-Gaussian case it receives connected higher-order contributions that we do not model here. These have been shown to be subdominant (e.g. \citealt{Colavincenzo18,Blot19})

For the galaxy auto-power spectrum we adopt
\begin{equation}
{\rm Cov}\!\left[P_{gg}(k_a),P_{gg}(k_b)\right]
=
\delta_{ab}\,
\frac{2\,P_{gg}^2(k_a)}{N_{\rm modes}(k_a)},
\end{equation}
where $N_{\rm modes}(k)$ is the total number of Fourier modes in the bin centred at $k$ (i.e. twice the number of independent modes).

For the galaxy--matter cross-power spectrum we use
\begin{equation}
{\rm Cov}\!\left[P_{gm}(k_a),P_{gm}(k_b)\right]
=
\delta_{ab}\,
\frac{P_{gg}(k_a)P_{mm}(k_a)+P_{gm}^2(k_a)}
{N_{\rm modes}(k_a)},
\end{equation}
and the Gaussian cross-covariance between $P_{gg}$ and $P_{gm}$ is
\begin{equation}
{\rm Cov}\!\left[P_{gg}(k_a),P_{gm}(k_b)\right]
=
\delta_{ab}\,
\frac{2\,P_{gg}(k_a)P_{gm}(k_a)}
{N_{\rm modes}(k_a)}.
\end{equation}

The diagonal covariance of the galaxy bispectrum is modelled as
\begin{equation}
\begin{aligned}
{\rm Cov}\!\left[B_{ggg}(t_a),B_{ggg}(t_b)\right]
= &\; \delta_{ab}\,
\frac{V}{N_{\triangle}(t_a)}\,
s_{123}\; \times \\ &\; P_{gg}(k_1)P_{gg}(k_2)P_{gg}(k_3),
\end{aligned}
\end{equation}
where $t_a\equiv(k_1,k_2,k_3)$ denotes a triangle bin, $V$ is the simulation volume, $N_{\triangle}$ is the number of fundamental triangles in the bin, and $s_{123}$ is the symmetry factor,
\begin{equation}
s_{123}=
\begin{cases}
6, & k_1=k_2=k_3,\\
2, & \text{two sides equal},\\
1, & \text{all sides different}.
\end{cases}
\end{equation}
In our implementation,
\begin{equation}
N_{\triangle}(k_1,k_2,k_3)
=
\frac{V^2}{(2\pi)^6}\,
8\pi^2\,k_1k_2k_3\,(\Delta k)^3,
\end{equation}
with $\Delta k$ the bispectrum bin width.

For the mixed bispectrum $B_{ggm}$ we analogously use
\begin{equation}
\begin{aligned}
{\rm Cov}\!\left[B_{ggm}(t_a),B_{ggm}(t_b)\right]
= &\;
\delta_{ab}\,
\frac{V}{N_{\triangle}(t_a)}\,
s_{123}\; \times \\ &\;
P_{gg}(k_1)P_{gg}(k_2)P_{mm}(k_3),
\end{aligned}
\end{equation}
and the Gaussian cross-covariance between $B_{ggg}$ and $B_{ggm}$, included in the joint fit, is
\begin{equation}
\begin{aligned}
{\rm Cov}\!\left[B_{ggg}(t_a),B_{ggm}(t_b)\right]
= &\;
\delta_{ab}\,
\frac{V}{N_{\triangle}(t_a)}\,
s_{123}\; \times \\ &\;
P_{gg}(k_1)P_{gg}(k_2)P_{gm}(k_3).
\end{aligned}
\end{equation}

Thus, for the joint fits to $\{P_{gg},P_{gm}\}$ and $\{B_{ggg},B_{ggm}\}$ we construct the block covariance matrices
\begin{equation}
\mathbf{C}_{P} =
\begin{pmatrix}
\mathbf{C}_{P_{gg}P_{gg}} & \mathbf{C}_{P_{gg}P_{gm}} \\
\mathbf{C}_{P_{gm}P_{gg}} & \mathbf{C}_{P_{gm}P_{gm}}
\end{pmatrix},
\end{equation}
and
\begin{equation}
\mathbf{C}_{B} =
\begin{pmatrix}
\mathbf{C}_{B_{ggg}B_{ggg}} & \mathbf{C}_{B_{ggg}B_{ggm}} \\
\mathbf{C}_{B_{ggm}B_{ggg}} & \mathbf{C}_{B_{ggm}B_{ggm}}
\end{pmatrix}.
\end{equation}
In the combined power-spectrum plus bispectrum fits, we neglect the cross-covariance between the power-spectrum and bispectrum sectors, so that the full covariance is taken to be block diagonal in $\mathbf{C}_P$ and $\mathbf{C}_B$.

Throughout, the Gaussian covariance is evaluated using the measured spectra of the catalogue under consideration. Contributions beyond the Gaussian approximation, such as connected trispectrum terms in the power-spectrum covariance, connected higher-order terms in the bispectrum covariance, and the non-Gaussian power-spectrum-bispectrum cross-covariance, are not included. These terms would generate additional covariance contributions, including off-diagonal correlations between bins. 

Moreover, although our measurements are averaged over paired-and-fixed simulations \citep{Angulo2016}, we evaluate the Gaussian covariance using the volume of a single simulation box. The covariance reduction specific to fixed-and-paired ensembles is non-trivial and is not modelled here \citep{Maion2022}.

The Gaussian approximation is not expected to be exact on the smallest scales considered in this work. However, our goal is not precision forecasting, but rather a controlled consistency test of model robustness. For that purpose, the analytic Gaussian covariance provides a transparent and computationally efficient approximation. A more complete covariance treatment, calibrated from a large ensemble of independent simulations and including non-Gaussian contributions, is left for future work.

\subsection{Likelihood and goodness of fit}

Given a model prediction $\mathbf{M}(\boldsymbol{\theta})$ for a parameter vector $\boldsymbol{\theta}$, we define the residual vector
$\boldsymbol{\Delta}(\boldsymbol{\theta}) \equiv \mathbf{D}-\mathbf{M}(\boldsymbol{\theta})$,
and the corresponding chi-squared as
\begin{equation}
\chi^2(\boldsymbol{\theta})
=
\boldsymbol{\Delta}^{\rm T}(\boldsymbol{\theta})\,
\mathbf{C}^{-1}\,
\boldsymbol{\Delta}(\boldsymbol{\theta}).
\end{equation}
For joint fits, the data vector, model vector, and covariance matrix are understood to contain the concatenated statistics described above.

Assuming a multivariate Gaussian likelihood, we write
\begin{equation}
\ln \mathcal{L}(\boldsymbol{\theta})
=
-\frac{1}{2}\chi^2(\boldsymbol{\theta})+{\rm const.}
\end{equation}
In practice, only differences in $\ln\mathcal{L}$ matter for inference, so the additive normalization constant is ignored. Since cosmology and the covariance are fixed in this work, the normalization is parameter-independent. It would need to be retained in analyses where the covariance varies with cosmology.

For each fit we monitor the minimum chi-squared and the reduced chi-squared as diagnostics of goodness of fit. However, as emphasised throughout this paper, goodness of fit alone is not used as the sole criterion for model validity. The behaviour of the recovered bias parameters is equally important.

\subsection{Priors and parameter sampling}

Posterior sampling is performed with the nested-sampling algorithm implemented in \texttt{MultiNest} \citep{multinest1,multinest2,multinest3}.

We adopt broad, uniform priors on all deterministic bias and stochastic nuisance parameters, chosen sufficiently wide that the posterior support is fully contained within the prior range. This ensures that the inferred constraints are driven by the data rather than by prior boundaries. The priors used in the analysis are
\begin{align}
b_1 &\in [-2,\,2], \nonumber\\
b_2 &\in [-4,\,4], \nonumber\\
b_{s^2} &\in [-5,\,5], \nonumber\\
b_{\nabla^2\delta} &\in [-40,\,40], \nonumber\\
b_3 &\in [-40,\,40], \nonumber\\
\epsilon_{P,1} &\in [-10,\,10], \nonumber\\
\epsilon_{P,2} &\in [-500,\,500], \nonumber\\
\epsilon_{B,i} &\in [-1000,\,1000], \qquad i=0,\dots,4, \nonumber\\
\eta_i &\in [-1000,\,1000], \qquad i=0,1,2.
\end{align}
We checked that these priors are uninformative and that the posteriors lie well within the prior range. For the fiducial second-order model, the parameter $b_3$ is of course absent.

In the runs presented here we use 250 live points and an evidence tolerance of 0.1. For each fit, the posterior distribution of all parameters is sampled, and summary statistics such as posterior means, standard deviations, and marginal credible intervals are extracted.

Nested sampling is particularly convenient in the present context because it efficiently explores multimodal or partially degenerate parameter spaces while simultaneously providing the Bayesian evidence, although the latter is not central to the present analysis.

\subsection{Model validation criteria}

The central test of this paper is performed by repeating the parameter inference with a fixed $k_{\max}^{P}$ while varying the maximum bispectrum scale $k_{\max}^{B}$.

Starting from conservative large-scale cuts, $k_{\max}^{B} = 0.1\,h\,{\rm Mpc}^{-1}$, progressively smaller triangle configurations are included in the data vector down to $k_{\max}^{B} = 0.4\,h\,{\rm Mpc}^{-1}$ in steps of $\Delta k_{\max}^{B} = 0.05\,h\,{\rm Mpc}^{-1}$. At each value of $k_{\max}^{B}$, we re-fit the model and track the posterior constraints on the deterministic bias parameters. This procedure allows us to test whether the Hybrid bias expansion remains stable as increasingly nonlinear information is introduced.

We interpret the model as valid only if it passes the following tests:
\begin{enumerate}
    \item \textit{Statistically acceptable description:} the fit provides a realistic description of the data.
    \item \textit{Consistency with independent calibration:} the posterior constraints remain statistically consistent with the probabilistic-bias measurements of Section~\ref{sec:prob_bias}.
    \item \textit{Stability under scale extension:} the inferred parameters do not show systematic drift as $k_{\max}^{B}$ is increased.
\end{enumerate}

The physically meaningful signal of failure is that the fitted clustering parameters cease to represent the same large-scale bias coefficients measured independently from the initial conditions. This criterion is intentionally conservative. Once the model requires scale-dependent or shifted bias parameters to absorb inaccuracies in the theoretical description, cosmological inference based on those fits would no longer be robust.

The maximum bispectrum scale quoted in this work is therefore defined as the largest value of $k_{\max}^{B}$ for which these consistency conditions remain satisfied.

\subsection{Perturbative EFT comparison}
\label{sec:pbj_method}

As a reference for the Hybrid model validation test, we also repeat the analysis using a standard perturbative EFT model implemented with the public \texttt{PBJ}\footnote{\url{https://chiaramoretti.gitlab.io/pbj/}} code \citep{Moretti_2023}. The purpose of this comparison is not to perform a fully optimised EFT analysis, but rather to assess, under the same simulation volume, tracer number densities, and covariance assumptions used in this work, the scale at which a perturbative description begins to show the same kind of bias-parameter instability used to define the Hybrid model breakdown scale.

As in the rest of this work, we keep the cosmology fixed to the simulation cosmology and fit only the galaxy auto-statistics, $\mathbf D_{\rm EFT}=\{P_{gg},B_{ggg}\}$ as implemented in \cite{Oddo_2020,Oddo_2021}.
We do not include $P_{gm}$ or $B_{ggm}$ in the perturbative-EFT comparison, since the \texttt{PBJ} setup used here does not provide the corresponding matter cross-statistics. The power-spectrum cut is fixed to $k_{\max}^P=0.3\,h\,{\rm Mpc}^{-1}$, which we found to be a conservative choice for the perturbative model in the present real-space setup, while the maximum bispectrum scale $k_{\max}^B$ is varied in the same way as in the Hybrid analysis.

The perturbative model consists of the one-loop real-space galaxy power spectrum and the tree-level real-space galaxy bispectrum. The power spectrum includes the Eulerian bias parameters $\{b^{\rm E}_1,\;b^{\rm E}_2,\;b^{\rm E}_{\mathcal G_2},\;b^{\rm E}_{\Gamma_3}\}$, together with a counterterm coefficient $c_0$ and stochastic parameters controlling the constant and scale-dependent shot-noise contributions. Schematically, the stochastic contribution is written as
\begin{equation}
P_{\rm noise}(k)=
\frac{1}{\bar n_g} \left(1+a_P+\epsilon_{k^2}k^2\right),
\end{equation}
The tree-level bispectrum depends on $b^{\rm E}_1$, $b^{\rm E}_2$, and $b^{\rm E}_{\mathcal G_2}$, and includes stochastic terms proportional to
$(P_{\rm L}(k_1)+P_{\rm L}(k_2)+P_{\rm L}(k_3))/\bar{n}$ and to $1/\bar{n}^2$, controlled by nuisance parameters $a_1$ and $a_2$. Here $P_{\rm L}$ refers to the linear power spectrum. Note that the one-loop bispectrum implementation could push its validity to smaller scales than the one explored here \citep{Angulo_2015,Eggmeier_2021,DAmico_2024b,Bakx_2025}. However, these come at the cost of including extra free parameters.

The full parameter vector sampled in the perturbative-EFT comparison is therefore
\begin{equation}
\boldsymbol{\theta}_{\rm EFT}
=
\{b_1,b_2,b_{\mathcal G_2},b_{\Gamma_3},c_0,a_P,\epsilon_{k^2},a_1,a_2\},
\end{equation}
for which we adopt broad uniform priors:
\begin{align}
b^{\rm E}_1 &\in [0.5,\,3.5], \nonumber\\
b^{\rm E}_2 &\in [-10,\,10], \nonumber\\
b^{\rm E}_{\mathcal G_2} &\in [-10,\,10], \nonumber\\
b^{\rm E}_{\Gamma_3} &\in [-10,\,10], \nonumber\\
c_0 &\in [-100,\,100], \nonumber\\
a_P &\in [-10,\,10], \nonumber\\
\epsilon_{k^2} &\in [-100,\,100], \nonumber\\
a_1 &\in [-10,\,10], \nonumber\\
a_2 &\in [-10,\,10].
\end{align}

The bias parameters returned by \texttt{PBJ} are defined in the Eulerian EFT basis, while the Hybrid model parameters and the probabilistic-bias measurements used throughout this paper are interpreted as Lagrangian bias coefficients. Before comparing the perturbative-EFT constraints to the probabilistic-bias benchmark, we therefore convert the relevant Eulerian parameters to the corresponding Lagrangian convention using the standard local-in-matter-density mapping \citep{Sheth_2013,Desjacques2018,Oddo_2021}:
\begin{align}
b_1 &= b_1^{\rm E}-1, \nonumber\\
b_2 &= b_2^{\rm E}-\frac{4}{3}b_{\mathcal G_2}^{\rm E}-\frac{8}{21}b_1, \label{eq:b2_EtoL}\\
b_{s^2} &= b_{\mathcal G_2}^{\rm E}+\frac{2}{7}b_1. \nonumber
\end{align}
Here $b_{\mathcal G_2}^{\rm E}$ denotes the Eulerian tidal-bias parameter used by the perturbative-EFT model. These transformed parameters,
$\{b_1^{\rm EFT},b_2^{\rm EFT},b_{s^2}^{\rm EFT}\}$,
are the quantities shown in the comparison plots and used in the Figure of Bias calculation. In practice, the transformation is applied directly to the posterior samples before computing means, covariances, and derived summary statistics. The remaining EFT parameters, such as $b_{\Gamma_3}$, counterterms, and stochastic amplitudes, are treated as nuisance parameters. We note that the perturbative-EFT comparison is less direct for the tidal-bias parameter than for $b_1$ and $b_2$. The PBJ model is written in an Eulerian Galileon basis, while the probabilistic-bias and Hybrid measurements use a Lagrangian tidal operator convention. This will become important in the next section.

For consistency with the Hybrid analysis, we monitor the transformed Lagrangian bias parameters as $k_{\max}^B$ is varied and use the same stability and probabilistic-bias consistency criteria to identify the perturbative-EFT breakdown scale.

\section{Results}
\label{sec:results}

\subsection{Best-fitting power spectrum and bispectrum}

\begin{figure}
    \includegraphics[width=\columnwidth]{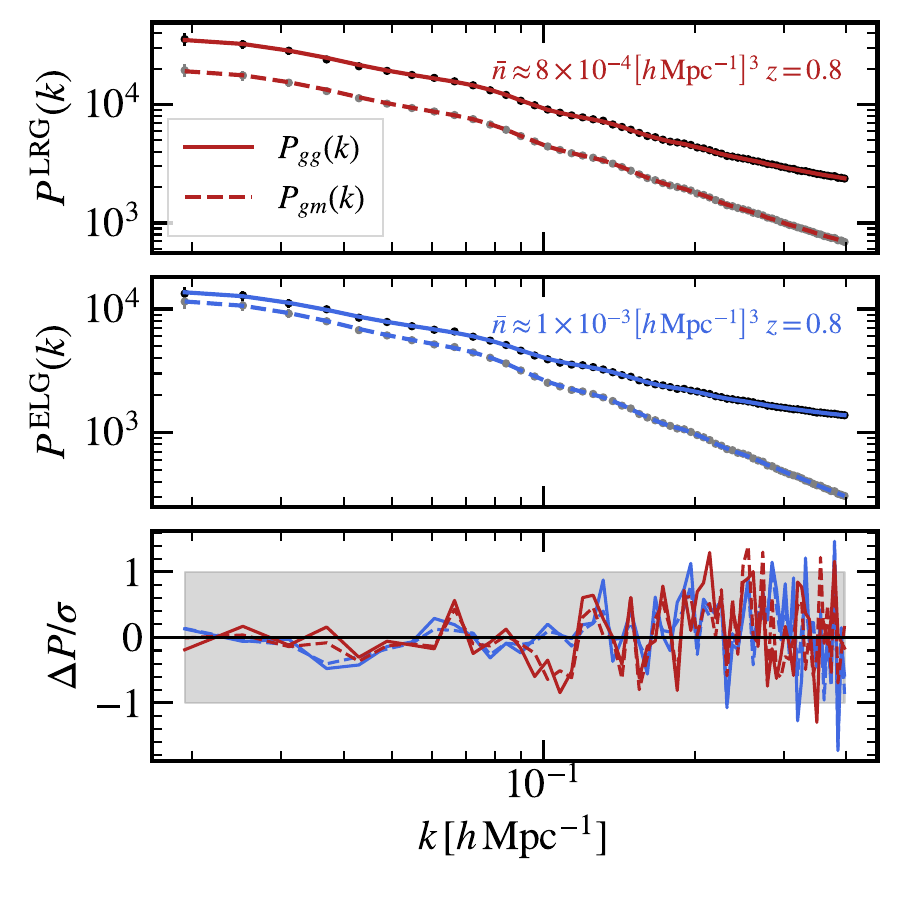}
    \caption{Best-fitting power spectra for the joint fit to power spectrum and bispectrum, using $k_{\rm max}^P=0.4\,h\,{\rm Mpc}^{-1}$ and $k_{\rm max}^B=0.25\,h\,{\rm Mpc}^{-1}$. The upper panels show the LRG and ELG auto- and matter-cross power spectra, while the lower panel shows the residuals in units of the adopted standard deviation.}
    \label{fig:power_fit}
\end{figure}

\begin{figure*}
    \includegraphics[width=\textwidth]{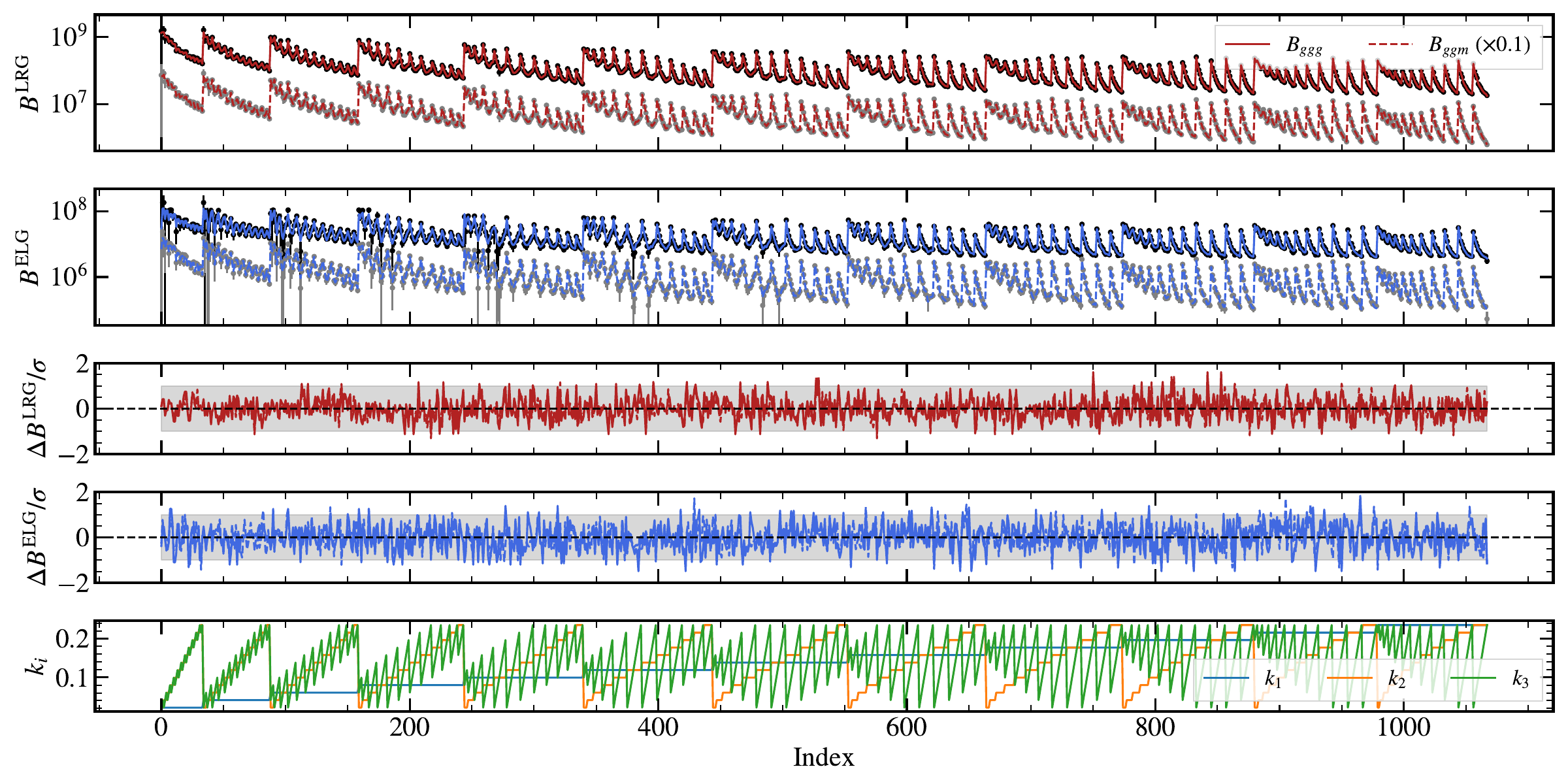}
    \caption{Best-fitting bispectra for the same joint fit as in Fig.~\ref{fig:power_fit}. The upper panels show the LRG and ELG measurements of $B_{ggg}$ and $B_{ggm}$, the middle panels show the corresponding residuals in units of the adopted standard deviation, and the lower panel indicates the ordered $(k_1,k_2,k_3)$ triangle configurations included in the fit.}
    \label{fig:bispec_fit}
\end{figure*}

We begin by illustrating the quality of the Hybrid model fits at the scale cuts that we ultimately identify as conservative and robust, namely
$k_{\rm max}^P=0.4\,h\,{\rm Mpc}^{-1}$ for the power spectrum and
$k_{\rm max}^B=0.25\,h\,{\rm Mpc}^{-1}$ for the bispectrum. The corresponding best-fitting models for the full data vector,
$P_{gg}(k)+P_{gm}(k)+B_{ggg}(k_1,k_2,k_3)+B_{ggm}(k_1,k_2,k_3)$,
are shown in Figs~\ref{fig:power_fit} and \ref{fig:bispec_fit}.

The agreement is excellent for both tracer samples. The power spectra are accurately reproduced over the full fitted range, and the bispectra show residuals consistent with the adopted covariance model, with no obvious systematic trends as a function of triangle configuration. For the power spectra, the residuals lie well within the nominal $1\sigma$ band on large scales and fluctuate around zero at higher $k$, as seen in Fig.~\ref{fig:power_fit}. The same is true for the bispectra in Fig.~\ref{fig:bispec_fit}, where both $B_{ggg}$ and $B_{ggm}$ are described accurately up to $k_{\rm max}^B=0.25\,h\,{\rm Mpc}^{-1}$ for both the LRG and ELG mocks.

Although these fits are visually and statistically accurate, they should be interpreted together with the parameter-stability tests presented below. As defined in Section~\ref{sec:methodology}, our validation criterion requires not only an acceptable fit to the spectra, but also stable bias parameters consistent with the probabilistic-bias benchmark. The scale cuts used in Figs.~\ref{fig:power_fit} and \ref{fig:bispec_fit} therefore correspond to the smallest scales for which both conditions are satisfied.

A final point concerns the overall amplitude of the residuals. For the LRG sample, the reduced chi-squared of the best-fitting joint model is $\chi^2_{\rm red,full}=0.373$, with a very similar value obtained for the ELG sample. This value is below unity. This is expected because the paired-and-fixed simulations suppress cosmic variance on large scales more strongly than assumed by our Gaussian covariance model \citep{Angulo2016,Maion2022}, which is evaluated as if the measurements came from a standard ensemble of independent realisations. As a simple illustration, if the effective covariance were uniformly smaller by a factor of two due to the paired simulations, the corresponding reduced chi-squared would increase by the same factor, since $\chi^2\propto \mathbf C^{-1}$. In that case, the value quoted above would shift to approximately $0.75$, and the fit would still be statistically acceptable. We stress, however, that this should be regarded only as a heuristic check, since the variance suppression from paired-and-fixed simulations is scale dependent and is not expected to rescale all covariance blocks in exactly the same way \citep{Maion2022}.

\subsection{Scale dependence of recovered bias parameters}

\begin{figure*}
    \includegraphics[width=\columnwidth]{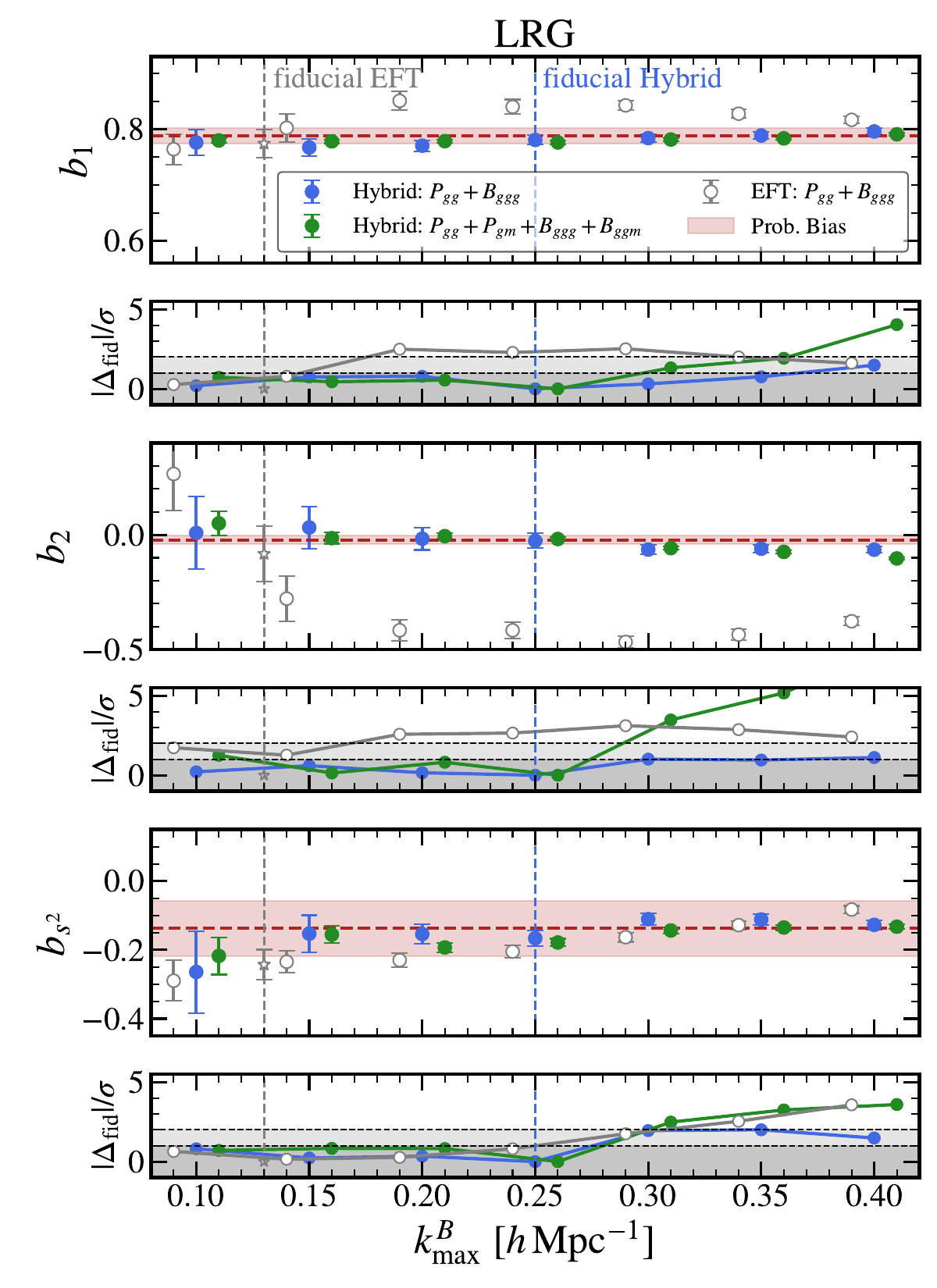}
    \includegraphics[width=\columnwidth]{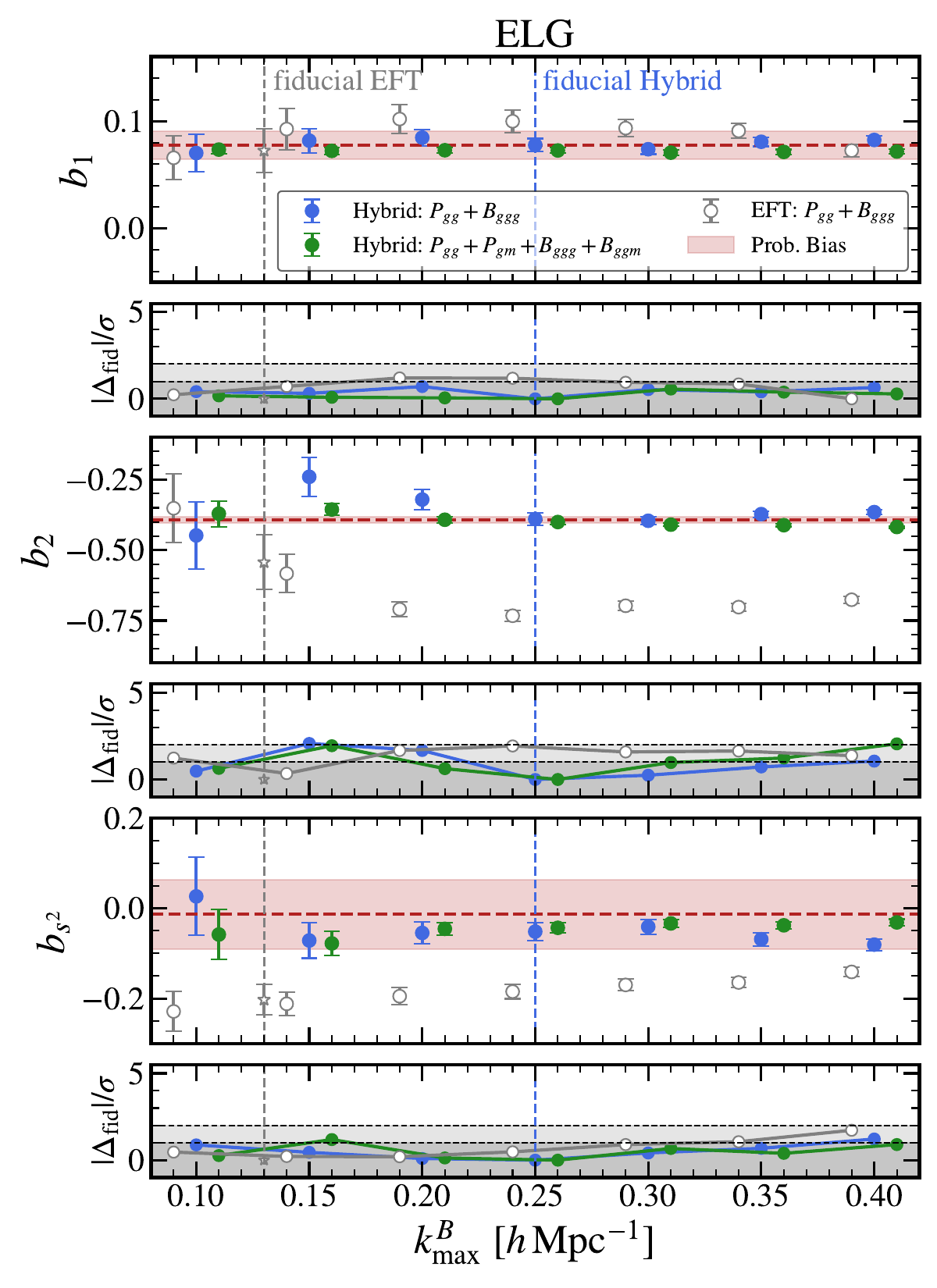}
    \caption{Marginalised bias constraints as a function of the maximum bispectrum scale. Left-hand panels show fits to the LRG sample, while right-hand panels show fits to the ELG sample for both the $P_{gg}+B_{ggg}$, and the $P_{gg}+P_{gm}+B_{ggg}+B_{ggm}$ cases. Larger panels show the absolute measurements by the Hybrid model. Smaller panels show the statistical differences with respect to the fiducial selected value corresponding to $k_{\rm max}^B=0.25\,h\,{\rm Mpc}^{-1}$ for the Hybrid approach and $k_{\rm max}^P=0.13\,h\,{\rm Mpc}^{-1}$ for the EFT approach. Note that $\Delta_{\rm fid}$ is a test of scale dependencies and does not compare the Probabilistic Bias measurements with the power plus bispectrum measurements. The grey shaded bands correspond to $1 \sigma$ and $2\sigma$ regions. The red shaded bands indicate the probabilistic-bias measurements adopted as the reference large-scale values. Thus, this plot shows both consistency with the independently measured probabilistic bias parameters and consistency within scales. We have shifted the EFT and cross spectra results to avoid clutter. }
    \label{fig:margestats_auto}
\end{figure*}

The central result of this paper is obtained by tracking the recovered bias parameters as a function of the maximum bispectrum scale. Figure~\ref{fig:margestats_auto} shows the one-dimensional marginalised constraints for the LRG and ELG samples. In each case, the left-hand panels correspond to fits using galaxy clustering only,
$P_{gg}+B_{ggg}$, while the right-hand panels include the matter cross-statistics,
$P_{gg}+P_{gm}+B_{ggg}+B_{ggm}$.

As expected, including the matter information breaks important degeneracies and substantially improves the precision of the recovered bias parameters. For this reason, the joint fits including $P_{gm}$ and $B_{ggm}$ provide our main reference for assessing model validity, while the galaxy-only fits are shown primarily to illustrate the gain in constraining power.

One feature of Fig.~\ref{fig:margestats_auto} is that, for some parameters, the uncertainty of the probabilistic-bias benchmark is larger than that obtained from the joint power-spectrum and bispectrum fit. This should not be interpreted as evidence that the clustering analysis provides a more fundamental measurement of the bias parameters. The two uncertainty estimates correspond to different statistical problems. The probabilistic-bias measurements are obtained at field level using the spatial-order-two tensorial estimators described in Section~\ref{sec:prob_bias}. In particular, the additional operators $J_{2=4}$ and $J_{4=4}$ are included in the estimator to reduce the smoothing-scale dependence of the lower-order bias parameters, even though they are not part of the fiducial basis fitted to the clustering statistics. Their inclusion propagates into the covariance of the lower-order parameters and can broaden the probabilistic-bias uncertainties. By contrast, the clustering constraints do not include these terms, an therefore, they are not marginalised over. Furthermore, these constraints use a Gaussian covariance approximation and neglect non-Gaussian mode coupling and power-spectrum--bispectrum cross-covariances. This approximation can make the clustering posteriors somewhat overconfident, especially on the smallest scales. A fully consistent comparison of the absolute error bars would require a covariance calibrated from a large ensemble of independent mock catalogues, which is beyond the scope of this work. Here we therefore focus primarily on the stability of the recovered parameters with scale and on their consistency with the probabilistic-bias benchmark.

Following the criterion defined in Section~\ref{sec:methodology}, we therefore look for two signatures of model breakdown. First, we test whether the recovered bias parameters remain stable as $k_{\rm max}^B$ is increased. If a parameter shifts systematically beyond its statistical uncertainty, the assumption of scale-independent bias coefficients is no longer valid. Second, we compare the fitted values to the independent probabilistic-bias measurements. If the clustering constraints cease to agree with the probabilistic-bias benchmark, then the fit is no longer recovering the physically meaningful large-scale bias parameters of the tracer sample. For clarity, we leave the Laplacian bias out of this comparison. Its value depends on the smoothing scale used to construct the basis fields, and although it can be adjusted by varying that scale, it remains close to zero in the present fits and does not affect the inferred breakdown scale.

For the LRG sample, the transition is clear. Up to
$k_{\rm max}^B\simeq 0.25\,h\,{\rm Mpc}^{-1}$,
the fitted parameters remain broadly stable and consistent with the probabilistic-bias values. Once triangles with
$k_{\rm max}^B\gtrsim 0.3\,h\,{\rm Mpc}^{-1}$
are included, both $b_2$ and $b_{s^2}$ begin to drift systematically, and the tension with the probabilistic-bias benchmark becomes significant. This is particularly evident in the joint fit including the matter cross-statistics, where the smaller posterior uncertainties make the transition sharper. We therefore interpret
$k_{\rm max}^B \simeq 0.25\,h\,{\rm Mpc}^{-1}$
as the largest bispectrum scale for which the Hybrid description remains self-consistent for the LRG-like sample.

The ELG sample shows a milder behaviour. The recovered parameters evolve more smoothly with $k_{\rm max}^B$, and there is no comparably sharp transition at a single scale. This is physically plausible, since ELGs are more weakly biased and trace the matter field more closely. At the same time, the ELG constraints do show mild tensions between the galaxy-only and galaxy-plus-matter analyses, as well as small offsets relative to the probabilistic-bias values at scales larger than $k_{\rm max}^B \simeq 0.25\,h\,{\rm Mpc}^{-1}$. We therefore do not regard the ELG results as compelling evidence that the model remains reliable beyond the LRG breakdown scale. Instead, we adopt the conservative conclusion that the Hybrid Bias Expansion model remains accurate down to
\begin{equation}
k_{\rm max}^B \simeq 0.25\,h\,{\rm Mpc}^{-1},
\end{equation}
while scales beyond this point show evidence that the assumptions of the model are beginning to fail.

In practical terms, it implies that the bispectrum-level Hybrid description extends the usable range of real-space clustering information substantially beyond the scale usually associated with standard perturbative EFT analyses, which typically break at around
$k_{\rm max}^B\simeq 0.15\,h\,{\rm Mpc}^{-1}$ \citep{DAmico_2024}.
The gain is therefore significant, but it is also finite: once the fit is pushed beyond
$k_{\rm max}^B\simeq 0.25\,h\,{\rm Mpc}^{-1}$,
the recovered bias coefficients no longer behave as true scale-independent large-scale parameters. This is in contrast with previous Hybrid studies of the power spectrum, where the model could be extended to smaller scales without comparably clear evidence of breakdown \citep{Modi_2020,ZennaroAnguloPellejero2021,PellejeroIbanez2022,PellejeroIbanez2023,PellejeroIbanez_2024}. At the same time, our probabilistic-bias measurements, obtained with the framework of \citet{StueckerPB_2024}, remain stable down to smaller smoothing scales, suggesting that the underlying bias description of large-scale structure should remain meaningful at least down to $k\sim 0.4\,h\,{\rm Mpc}^{-1}$, but not necessarily in the form of the low-order polynomial bias expansion adopted here. The inclusion of higher derivative terms $J_{2=4}$ and $J_{4=4}$ become relevant on these smaller scales while the approximation of the bias function as a polynomial can impact results (for more details, see \citealt{StueckerGPB_2024}).

\subsection{Comparison with perturbative EFT}
\label{sec:eft_comparison}

To put the Hybrid model breakdown scale in context, we repeat the same validation exercise using the perturbative EFT model described in Section~\ref{sec:pbj_method}. This comparison is performed on the same mock catalogues and with the same covariance prescription as the Hybrid analysis, but using only the galaxy auto-statistics, $P_{gg}+B_{ggg}$, since the \texttt{PBJ} implementation used here does not include the matter cross-statistics $P_{gm}$ and $B_{ggm}$. The cosmology is fixed to the simulation cosmology, and the perturbative-EFT bias parameters are allowed to vary. We fix the power-spectrum cut to $k_{\rm max}^P=0.3\,h\,{\rm Mpc}^{-1}$ and vary the maximum bispectrum scale $k_{\rm max}^B$.

The resulting constraints are shown together with the Hybrid constraints in Fig.~\ref{fig:margestats_auto}. For the perturbative EFT case, the plotted parameters correspond to the Eulerian PBJ bias parameters transformed to the Lagrangian convention, as described in Section~\ref{sec:pbj_method}. This allows a direct comparison with the probabilistic-bias benchmark used throughout this work.

For both tracer samples, the perturbative-EFT constraints show evidence of scale dependence at substantially larger scales than the Hybrid constraints. In the LRG sample, the effect is especially clear in $b_1$ and $b_2$: once bispectrum configurations beyond the largest scales are included, the recovered parameters drift away from the probabilistic-bias values and from the stable large-scale trend. A finer scan in $k_{\rm max}^B$ shows that the onset of this instability occurs at approximately $k_{\rm max}^B \simeq 0.13\,h\,{\rm Mpc}^{-1}$, for the tree-level bispectrum model (highlighted in Fig. \ref{fig:margestats_auto} as a star symbol). In the LRG case, all bias parameters are in broad agreement with the Probabilistic Bias measurements.

The ELG sample exhibits the same qualitative behaviour, although the detailed parameter shifts differ because of the different bias values of the tracer. As in the LRG case, the perturbative-EFT fit becomes unstable before the Hybrid fit does. This confirms that the comparison is not driven by a peculiarity of a single tracer sample, but reflects the more limited range of validity of the tree-level perturbative bispectrum model under the assumptions of this analysis.

We note that the perturbative-EFT comparison is particularly sensitive to the tidal-bias convention. The PBJ model is written in an Eulerian Galileon basis, which we transform to the Lagrangian $\{b_1,b_2,b_{s^2}\}$ convention before comparison. In this transformation we account for the relation between the Galileon operator and the traceless tidal operator, given in Eqs~\ref{eq:b2_EtoL}. The residual offset observed in the ELG tidal parameter is therefore unlikely to be driven solely by the Eulerian--Lagrangian mapping. More plausibly, it reflects the fact that in the tree-level perturbative bispectrum the tidal-bias direction can absorb missing higher-order or stochastic contributions, especially for a weakly biased tracer and in the absence of matter cross-statistics. We therefore interpret the EFT comparison primarily as a scale-stability diagnostic rather than as a precision measurement of each individual bias coefficient.

This test is not intended as a fully optimised perturbative-EFT analysis. In particular, we use the tree-level bispectrum, keep the cosmology fixed, and do not include matter cross-statistics. Nevertheless, it provides a useful reference: for the same simulation volume, number densities, and covariance model, the perturbative-EFT bispectrum begins to show bias-parameter scale dependence around $k_{\rm max}^B\simeq 0.13\,h\,{\rm Mpc}^{-1}$, while the Hybrid model remains self-consistent up to $k_{\rm max}^B\simeq 0.25\,h\,{\rm Mpc}^{-1}$. This comparison illustrates the gain provided by the hybrid treatment of nonlinear gravitational evolution.

\subsection{Figure of Bias and Figure of Merit}

We now quantify the previous conclusions using two summary statistics in the three-dimensional parameter subspace spanned by
$\boldsymbol{b}\equiv (b_1,b_2,b_{s^2})$.
As before, we leave the Laplacian bias out of this comparison because of its explicit dependence on the basis smoothing scale.

For each value of $k_{\rm max}^B$, we define the difference between the clustering-inferred mean parameters and the probabilistic-bias benchmark as
$\Delta \boldsymbol{b} = \bar{\boldsymbol{b}}_{\rm fit} - \bar{\boldsymbol{b}}_{\rm PB}$, and the corresponding total covariance as $\mathbf C_{\rm tot} = \mathbf C_{\rm fit} + \mathbf C_{\rm PB}$,
where $\mathbf C_{\rm fit}$ is the posterior covariance matrix from the clustering analysis and $\mathbf C_{\rm PB}$ is the covariance of the probabilistic-bias measurements.

The Figure of Bias (FoB) is then defined from the distance between the two parameter vectors,
\begin{equation}
{\rm FoB} = \frac{\sqrt{\Delta \boldsymbol{b}^{\rm T}\mathbf C_{\rm tot}^{-1}\Delta \boldsymbol{b}}}{\sqrt{\chi^2_{\rm lim}}},
\end{equation}
with $\chi^2_{\rm lim} = \chi^2_{\nu=3}\!\left(0.68\right)$,
where $\chi^2_{\nu=3}(p)$ denotes the $p$-quantile of the $\smash{\chi^2}$ distribution with three degrees of freedom. Note that $\nu=3$ refers to the three degrees of freedom on which we focus, $\{b_1,b_2,b_{s^2}\}$, since we have left the Laplacian out of the comparison due to its dependencies with the smoothing scales. With this normalization, ${\rm FoB}=1$ corresponds to the chosen 68\% reference in the three-dimensional bias-parameter space.

\begin{figure}
    \includegraphics[width=\columnwidth]{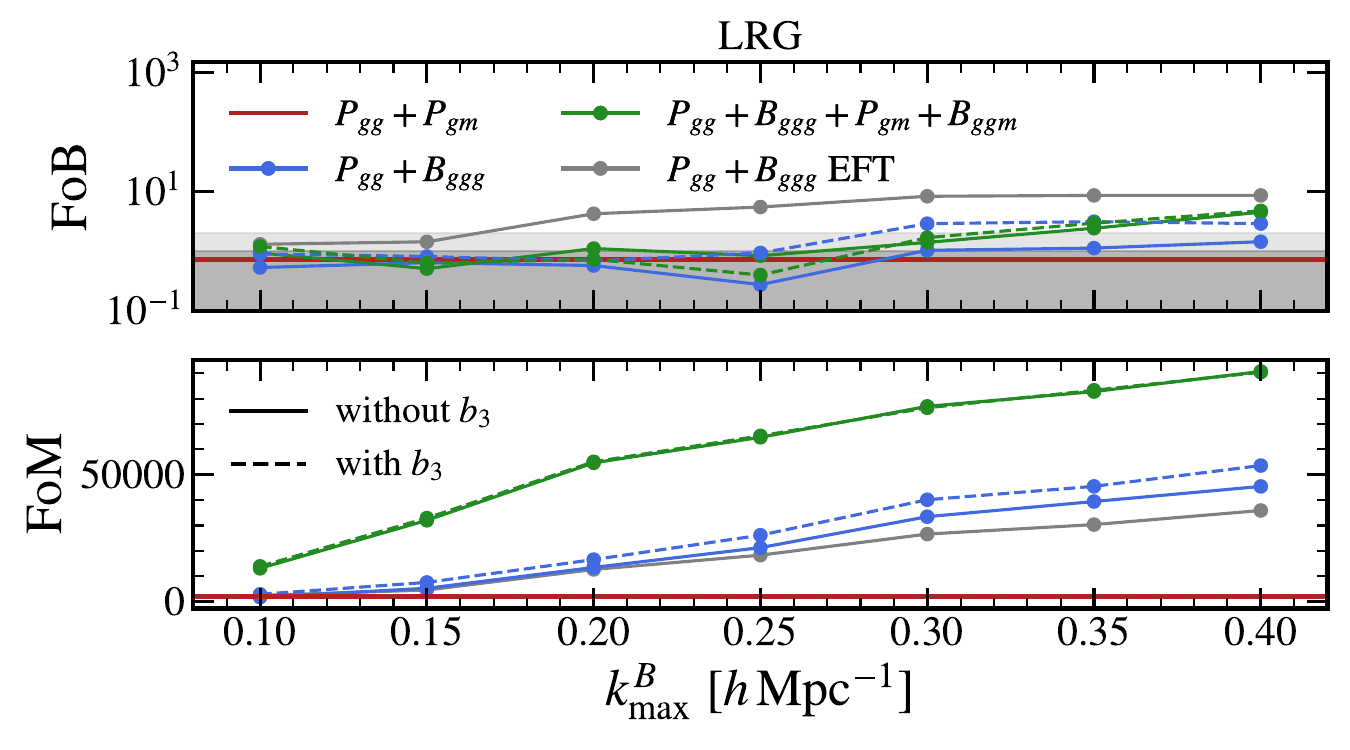}
    \includegraphics[width=\columnwidth]{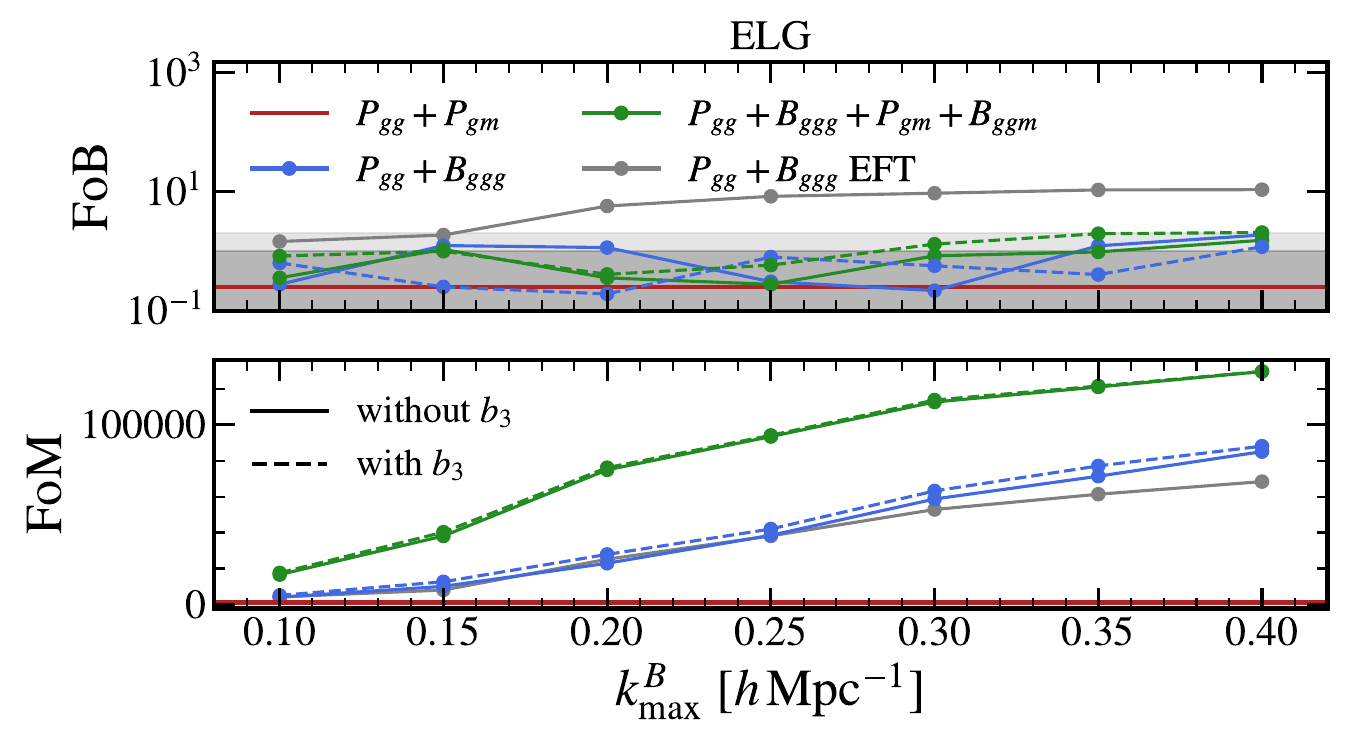}
    \caption{Figure of Bias and Figure of Merit as a function of the maximum bispectrum scale for the LRG and ELG samples. Solid coloured lines correspond to the fiducial second-order Hybrid model, while dashed coloured lines show the partial third-order extension discussed in Section~\ref{sec:third_order}. The red horizontal line shows the Hybrid power-spectrum-only reference fit, $P_{gg}+P_{gm}$, evaluated at $k_{\rm max}^P=0.4\,h\,{\rm Mpc}^{-1}$. The grey curve shows the perturbative-EFT comparison based on $P_{gg}+B_{ggg}$, with $k_{\rm max}^P=0.3\,h\,{\rm Mpc}^{-1}$. Shaded bands represent the 1-$\sigma$ and 2-$\sigma$ regions.}
    \label{fig:FOMB}
\end{figure}

The Figure of Merit (FoM) is defined as
\begin{equation}
{\rm FoM} = \frac{1}{\sqrt{\det \mathbf C_{\rm tot}}},
\end{equation}
which is inversely proportional to the volume of the allowed region in this parameter subspace. Larger FoM values therefore correspond to tighter joint constraints.
The results are shown in Fig.~\ref{fig:FOMB}.

In addition to the fits including the bispectrum, Fig.~\ref{fig:FOMB} shows as a red horizontal reference line the result obtained from the power-spectrum-only fit, $P_{gg}+P_{gm}$, evaluated at the same power-spectrum cut used in the rest of the Hybrid analysis, $k_{\rm max}^P=0.4\,h\,{\rm Mpc}^{-1}$. This provides a clean baseline for isolating the information added by the bispectrum as $k_{\rm max}^B$ is increased. The power-spectrum-only FoM is substantially smaller than the FoM obtained once bispectrum information is included. The fact that it remains well below the bispectrum-including constraints shows that the bispectrum provides substantial independent information on the bias sector, rather than merely acting as a consistency check of the power-spectrum fit.

The grey curves show the perturbative-EFT comparison described in Section~\ref{sec:eft_comparison}, based on fits to $P_{gg}+B_{ggg}$ with $k_{\rm max}^P=0.3\,h\,{\rm Mpc}^{-1}$. These curves should be interpreted as a validation test rather than as a fully optimised EFT likelihood. They show that, under the same mock-data and covariance assumptions, the perturbative tree-level bispectrum becomes inconsistent with the probabilistic-bias benchmark at substantially larger physical scales, i.e. lower $k_{\rm max}^B$, than the Hybrid model.

For the LRG sample, the FoB confirms the picture obtained from the parameter-evolution plots. The Hybrid model tension remains modest up to
$k_{\rm max}^B\simeq 0.25\,h\,{\rm Mpc}^{-1}$,
and then increases rapidly between
$k_{\rm max}^B\simeq 0.25$
and
$0.3\,h\,{\rm Mpc}^{-1}$.
This behaviour is seen both in the galaxy-only analysis and in the joint analysis including matter cross-statistics, with the latter giving the sharper and more significant transition. By contrast, the perturbative-EFT FoB rises much earlier, already becoming large once the bispectrum fit is pushed beyond the largest scales. This provides a compact quantitative confirmation that the EFT tree-level bispectrum breaks down around
$k_{\rm max}^B\simeq 0.13$--$0.15\,h\,{\rm Mpc}^{-1}$,
whereas the Hybrid description remains self-consistent up to approximately
$k_{\rm max}^B\simeq 0.25\,h\,{\rm Mpc}^{-1}$.

For the ELG sample, the FoB again shows a weaker trend for the Hybrid case. Most measurements remain within roughly the $2\sigma$ level, with no abrupt transition of the kind seen for LRGs. This is fully consistent with the smoother behaviour of the ELG posteriors in Fig.~\ref{fig:margestats_auto}. The perturbative-EFT comparison, however, again shows a much earlier increase in FoB than the Hybrid fits. We therefore regard the ELG case as broadly compatible with the same conservative Hybrid threshold, while the EFT comparison supports the conclusion that the perturbative tree-level bispectrum has a substantially shorter range of validity in this setup.

The FoM shows a complementary aspect of the analysis. For both tracer samples, the Hybrid model constraining power increases steadily as smaller bispectrum scales are included. In particular, extending the fit from
$k_{\rm max}^B\simeq 0.15$
to
$0.25\,h\,{\rm Mpc}^{-1}$
reduces the allowed parameter volume by roughly a factor of $2$--$3$ within the bispectrum-including analyses. Relative to the power-spectrum-only reference, the gain is much larger, reaching orders of magnitude in FoM. The perturbative-EFT FoM agrees well with the Hybrid bispectrum FoM over most of the range shown. However, the earlier loss of bias consistency in the perturbative model makes these constraints difficult to interpret.

Of course, the FoM shown here refers only to the bias-parameter subspace and does not translate directly into a cosmological FoM. Different cosmological parameters project onto the clustering observables in different ways. Nevertheless, the tightening of the bias constraints is still relevant for cosmological inference, since it helps to break degeneracies between nuisance and cosmological parameters, for example those involving combinations such as $b_1$ and $b_2$. In that sense, the gain in constraining power between $k_{\rm max}^B\simeq 0.15$ and $0.25\,h\,{\rm Mpc}^{-1}$ is likely to propagate at least partially into tighter cosmological constraints, provided the model remains valid over that range.

\subsection{Impact of a partial third-order extension}
\label{sec:third_order}

As a limited robustness test, we extend the deterministic bias expansion by including the cubic density operator $\delta^3$, with associated bias coefficient $b_3$. This increases the number of bispectrum basis terms from 125 in the second-order model to 216. For simplicity, this test is performed using only one member of the paired-and-fixed pair simulations, so the corresponding measurements are noisier than in the fiducial analysis, but the comparison is internally consistent.

The results are shown as dashed lines in Fig.~\ref{fig:FOMB}. The main conclusion is unchanged: adding $\delta^3$ does not extend the scale at which the description remains self-consistent. The recovered parameters show the same qualitative drifts with $k_{\rm max}^B$ and the same tensions with the probabilistic-bias benchmark as in the fiducial second-order analysis. In particular, the LRG transition around $k_{\rm max}^B\simeq 0.25$--$0.3\,h\,{\rm Mpc}^{-1}$ remains clearly visible.

This result should not be over-interpreted. The operator $\delta^3$ does not constitute a complete third-order basis: tidal cubic terms and higher-derivative cubic operators are omitted, and a full third-order treatment would require a much larger basis and a substantial increase in the number of bias parameters \citep{Desjacques2018,lazeyras_2019,Schmidt2021}. The present test is therefore only a limited check of whether the dominant cubic density contribution alone is sufficient to delay the onset of breakdown.

A minor subtlety is that, in the galaxy-only case, the projected FoM in the $\{b_1,b_2,b_{s^2}\}$ subspace can increase slightly when $b_3$ is added, especially at small $k_{\rm max}^B$. This does not indicate a genuine gain in total constraining power. Rather, it reflects a rotation of the posterior degeneracy directions after the extra parameter is introduced, which can reduce the projected volume of a lower-dimensional subspace even though the full parameter-space volume necessarily grows. We verified explicitly that the full higher-dimensional FoM behaves as expected. Thus, the partial third-order extension changes the geometry of the posterior, but not the basic physical conclusion: the Hybrid model still breaks down once the bispectrum fit is pushed beyond
$k_{\rm max}^B\simeq 0.25\,h\,{\rm Mpc}^{-1}$.

\section{Relative importance of bispectrum basis terms}
\label{sec:basis_terms}

We now examine which groups of Hybrid bispectrum basis terms are most relevant in practice. This question is important beyond the scope of the present paper, since future efforts to build emulators for the Hybrid basis terms will probably be limited by the cost of measuring a large number of bispectra from simulations \citep{Angulo_2021,ZennaroAnguloPellejero2021,baccoemu2023,PellejeroIbanez_2024}. If only a subset of terms contributes significantly to the final model, emulator efforts can be focused on those terms, while subdominant contributions may be treated at lower precision or with simpler approximations.

Rather than ranking every individual bispectrum contribution, we group the basis terms by their total bias-operator order. We assign order zero to the constant operator, order one to the linear-density operator, and order two to the quadratic and higher-derivative operators,
${\rm ord}(1)=0$,
${\rm ord}(\delta)=1$, and
${\rm ord}(\delta^2)={\rm ord}(s^2)={\rm ord}(\nabla^2\delta)=2$.
In the partial third-order extension, the operator $\delta^3$ is assigned order three. The order of a bispectrum basis term is then defined as the sum of the orders of the three operators entering the correlator. For example,
$\langle 1\,1\,1\rangle$ is zeroth order,
$\langle 1\,1\,\delta\rangle$ and its permutations are first order,
$\langle 1\,1\,\delta^2\rangle$, $\langle 1\,1\,s^2\rangle$, $\langle 1\,1\,\nabla^2\delta\rangle$, and $\langle 1\,\delta\,\delta\rangle$ are second order, while $\langle \delta\,\delta\,\delta\rangle$ and $\langle 1\,\delta\,\delta^2\rangle$ are third order. Since the bispectrum contains three fields, the operator combinations considered here extend up to ninth order.

To quantify the contribution of each order, we compute the summed absolute amplitude of all terms belonging to that order and normalise by the total summed absolute amplitude over all terms,
\begin{equation}
f_n =
\frac{
\sum_{X\in n}\sum_{\rm triangles}
\left| B_X(k_1,k_2,k_3)\right|
}{
\sum_X\sum_{\rm triangles}
\left| B_X(k_1,k_2,k_3)\right|
}.
\end{equation}
For the bias-weighted cases, each basis bispectrum is first multiplied by the corresponding product of bias coefficients before performing the sum. This gives a direct estimate of the fraction of the total model amplitude associated with each perturbative order for a given tracer.

\begin{figure}
    \includegraphics[width=\columnwidth]{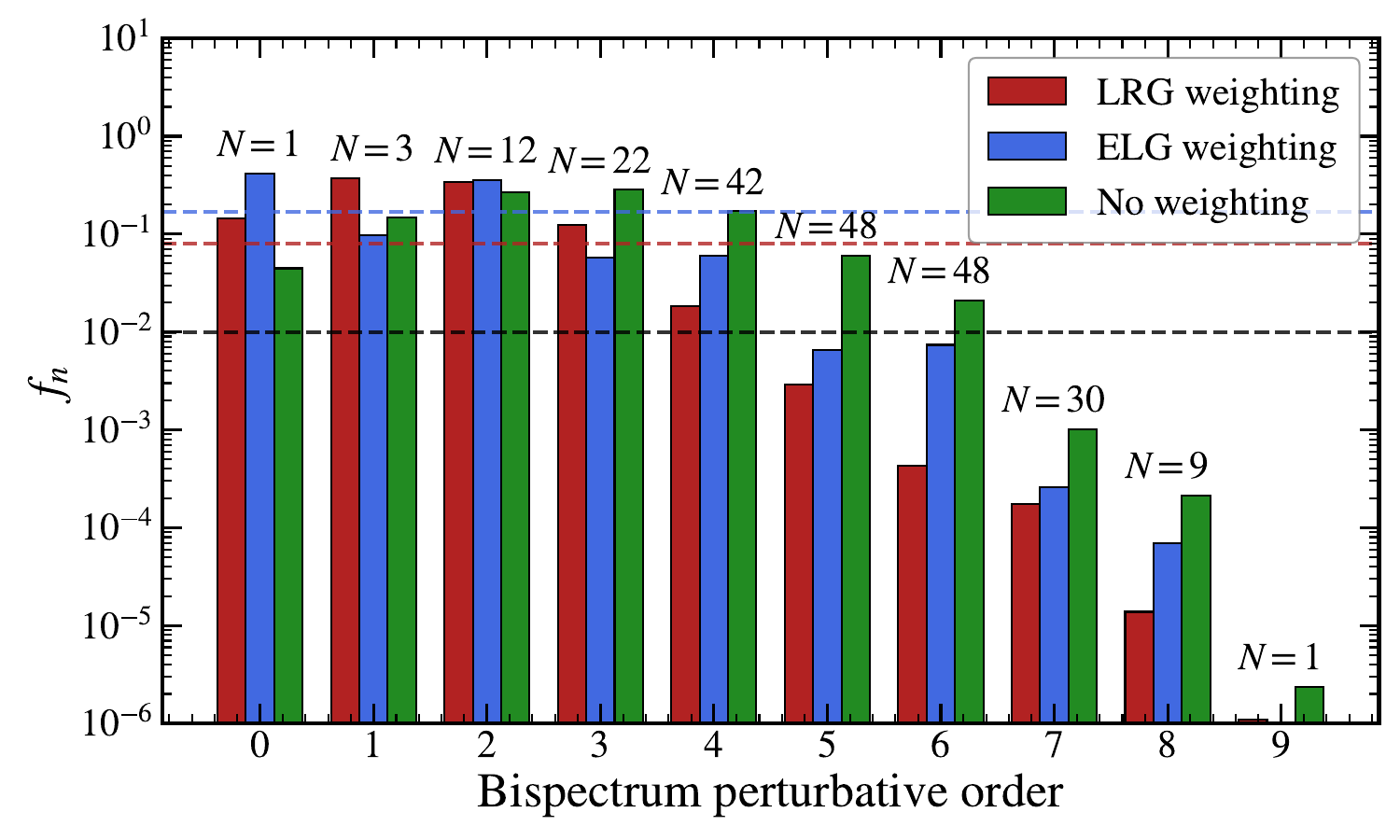}
    \caption{Fraction of the total summed bispectrum amplitude contributed by each bias-operator order. Bars show the result after weighting by representative LRG and ELG bias parameters, together with the unweighted basis hierarchy. The labels indicate the number of basis terms contributing to each order in the operator set considered here. The black dashed line marks a $1\%$ reference threshold, while the red and blue dashed lines indicate the approximate fractional precision expected for DESI-like LRG2 and ELG1 samples under the Gaussian covariance approximation at scales of $k_{\rm max}^B=0.25\,h\,{\rm Mpc}^{-1}$. Since the partial third-order extension includes only the $\delta^3$ operator, the term counts from fourth order onward do not represent the complete basis of a full third-order expansion.}
    \label{fig:amplitudes}
\end{figure}

The results are shown in Fig.~\ref{fig:amplitudes}. The bars show the amplitude fractions for representative LRG and ELG bias weightings, as well as the unweighted case. The labels above the bars indicate the number of basis terms contributing to each order. The black dashed line marks a conservative $1\%$ reference threshold. The red and blue dashed lines indicate the approximate fractional precision expected for DESI-like LRG2 and ELG1 samples, respectively. These thresholds provide a practical guide for deciding which operator orders must be modelled accurately for a given survey sample.

The expected hierarchy with perturbative order is clearly visible. The lowest-order terms carry most of the total amplitude, while the contribution decreases rapidly at high order. For both LRG and ELG weightings, the bulk of the signal is contained in the first few orders, with the precise distribution depending on the tracer bias parameters. The ELG weighting gives relatively more importance to second-order contributions, reflecting the larger role of quadratic-density terms for this tracer. The unweighted case provides a useful tracer-independent view of the intrinsic basis hierarchy, showing that the decrease with order is not solely produced by the numerical values of the LRG or ELG bias parameters.

The practical implication is that future Hybrid bispectrum emulators should prioritise the accurate prediction of the low-order basis sectors. If no specific tracer sample is assumed, the $1\%$ threshold provides a conservative criterion for deciding which orders should be included in a high-accuracy emulator. For a specific analysis, however, the relevant threshold should be set by the statistical precision of the sample. In that case, the DESI-like LRG2 and ELG1 reference lines in Fig.~\ref{fig:amplitudes} indicate that only orders contributing above the corresponding fractional error are likely to require the most accurate modelling.

A caveat is that our partial third-order extension includes only the density operator $\delta^3$, rather than the complete set of third-order operators. Consequently, from fourth order onward the number of terms shown in Fig.~\ref{fig:amplitudes} does not correspond to the complete basis that would arise in a full third-order bias expansion. The comparison among the highest orders is therefore not fully exhaustive. Nevertheless, the amplitudes of these higher-order sectors are already strongly suppressed in the basis considered here, so we do not expect the qualitative hierarchy to change substantially once the remaining third-order operators are included. A complete third-order treatment would be required to make this statement fully quantitative.

This analysis suggests that the computational cost of bispectrum emulation can be reduced by exploiting the strong hierarchy among basis contributions. The conclusion is not that high-order terms are universally negligible, but rather that their required modelling accuracy should be matched to the statistical precision of the target sample.

\section{Conclusions}
\label{sec:conclusion}

In this work we have investigated the range of validity of the Hybrid Bias Expansion (also known as HEFT) description of galaxy clustering at the bispectrum level in real space. Using DESI-like LRG and ELG mock catalogues built at fixed cosmology, we tested whether the Hybrid Bias Expansion can recover scale-independent Lagrangian bias parameters when fitted jointly to the power spectrum and bispectrum.

Our central result is that the bispectrum-level Hybrid model remains self-consistent up to $k_{\rm max}^B \simeq 0.25\,h\,{\rm Mpc}^{-1}$,
while clear signs of breakdown appear once triangles with
$k_{\rm max}^B \gtrsim 0.3\,h\,{\rm Mpc}^{-1}$
are included. For the LRG sample, this transition is particularly clear: the recovered bias parameters begin to drift systematically with scale and show significant tension with the independent probabilistic-bias measurements. The ELG sample shows a smoother behaviour, but does not provide compelling evidence that the model can be extended safely beyond the same conservative threshold.

A key aspect of this analysis is that model validity is not assessed purely through goodness of fit. The model is regarded as reliable only as long as the fitted clustering parameters remain consistent with the independently measured large-scale bias coefficients obtained from the probabilistic-bias framework, and remain stable as the maximum bispectrum scale is varied. This provides a physically motivated and conservative breakdown criterion: once the fitted parameters cease to behave as true scale-independent biases, the model can no longer be trusted for robust cosmological inference, even if the spectra themselves are still fitted accurately.

The resulting validity scale represents a substantial improvement over the range usually associated with standard perturbative EFT treatments of the bispectrum, which breaks down around
$k_{\rm max}^B \sim 0.13\,h\,{\rm Mpc}^{-1}$, as shown in our tailored test. A Hybrid Bias Expansion therefore makes it possible to exploit a significantly larger fraction of the mildly non-linear bispectrum information while still maintaining control over the physical interpretation of the bias parameters.

We also explored several secondary aspects of the problem. First, adding the matter cross-statistics, $P_{gm}$ and $B_{ggm}$, substantially improves the precision of the recovered bias parameters by breaking degeneracies present in galaxy-only clustering fits. Second, a partial third-order extension including only the $\delta^3$ operator does not shift the breakdown scale, indicating that the onset of failure is not cured by this minimal cubic correction alone.
Third, when the bispectrum basis terms are grouped by total bias-operator order, their summed amplitudes show a clear hierarchy: the lowest-order sectors dominate the signal, while higher-order contributions decrease rapidly. This suggests that future emulator efforts may be able to prioritise the most important operator-order sectors, with the required accuracy of subdominant sectors set by the statistical precision of the target sample.

The present work should be viewed as a controlled validation study rather than a complete end-to-end Hybrid model likelihood analysis. We have restricted attention to fixed cosmology and real space, and we have not attempted to construct a cosmology-dependent emulator for the full bispectrum basis. Natural next steps include extending the deterministic basis to a complete third-order operator set, quantifying the cosmology dependence of the dominant bispectrum basis terms, incorporating redshift-space distortions, and propagating the present findings into full cosmological parameter inference. These developments will be necessary to determine to what extent the gain in validity scale demonstrated here translates into improved cosmological constraints in realistic survey analyses.

\section*{Acknowledgements}

LL is supported by the Austrian Science Fund (FWF) [ESP 357-N]. This research was funded in part by the Austrian Science Fund (FWF) 10.55776/F101300. SC acknowledges the support of the `Ram\'on y Cajal' fellowship (RYC2023-043783-I). SC also acknowledges the support of the `Ayudas para Atracci\'on de Investigadores con Alto Potencial' (2025/00000640) from Universidad de Sevilla. MZ acknowledges that the project that gave rise to these results received the support of a fellowship from the `La Caixa' Foundation (ID 100010434); the fellowship code is LCF/BQ/PI25/12100027. JS acknowledges funding by the Austrian Science Fund (FWF) [10.55776/ESP705]
%
\section*{Data Availability}

 The data underlying this article will be shared on reasonable request to the corresponding author.



\bibliographystyle{mnras}
\bibliography{example} 





\bsp	
\label{lastpage}
\end{document}